\documentclass[journal]{new-aiaa}
\usepackage[utf8]{inputenc}
\usepackage{textcomp}

\usepackage{graphicx}
\usepackage{amsmath}
\usepackage[version=4]{mhchem}
\usepackage{siunitx}
\usepackage{longtable,tabularx}
\usepackage{siunitx}
\usepackage{makecell}
\usepackage{caption}
\usepackage{subcaption}
\usepackage{bm}
\usepackage{cuted}
\usepackage{mathtools}
\usepackage[extdef]{delimset}

\allowdisplaybreaks[1]

\newcommand{\vect}[1]{{\bm{#1}}}

\usepackage{amsthm}
\newtheorem{problem}{Problem}
\newtheorem{remark}{Remark}

\usepackage[ruled]{algorithm2e}
\usepackage{cleveref}

\crefname{equation}{Eq.}{Eqs.}
\Crefname{equation}{Equation}{Equations}

\crefname{figure}{Fig.}{Figs.}
\Crefname{figure}{Figure}{Figures}

\crefname{table}{Table}{Tables}
\crefname{section}{Section}{Sections}
\crefname{problem}{Problem}{Problems}
\crefname{algorithm}{Algorithm}{Algorithms}
\crefname{proposition}{Proposition}{Propositions}
\crefname{theorem}{Theorem}{Theorems}

\usepackage{multirow}
\title{Trajectory Dispersion Control for Precision Landing Guidance of Reusable Rockets}

\author{
Xinglun Chen \footnote{Ph.D. Candidate, School of Astronautics; chenxinglun@buaa.edu.cn.},
Ran Zhang \footnote{Associate Professor, School of Astronautics; zhangran@buaa.edu.cn (Corresponding Author).},
Huifeng Li \footnote{Professor, School of Astronautics; lihuifeng@buaa.edu.cn.}
}
\affil{Beihang University, 102206 Beijing, People's Republic of China}

\begin{document}

\maketitle

%% main text
\section{Introduction}
%\label{sec:introduction}

Precision landing guidance is a critical enabling technology for reusable rocket recovery.
Compared with lunar landing \cite{cherry1964general, klumpp1974apollo} and planetary landing \cite{acikmese2007convex, Lee2017Constrained}, 
endoatmospheric landing is subjected to more disturbing conditions, including engine thrust fluctuation, aerodynamic coefficient uncertainty, 
atmospheric density perturbation, and wind disturbance.
Under the effect of these disturbances, the rocket's flight state will exhibit uncertain variations, resulting in the presence of trajectory dispersion.
The trajectory dispersion, which can be characterized using the mean and variance of the trajectories, propagates with flight time and ultimately determines landing accuracy.
Therefore, this note focuses on the trajectory dispersion control problem, which directly optimizes the trajectory dispersions of both states and commands in real time, achieving high-precision landing of reusable rockets.

In the field of landing guidance methods, there are two major categories: deterministic guidance methods and probabilistic guidance methods.
The deterministic guidance methods, mainly including explicit guidance, trajectory tracking guidance, and Model Predictive Control (MPC) guidance, reduce the trajectory dispersion by attenuating the adverse effects of disturbances, which is an indirect trajectory dispersion control approach.
The explicit guidance method attenuates the effect of disturbances by regenerating feasible and/or optimal trajectories at each guidance period.
The representative explicit guidance methods mainly contain E-guidance \cite{cherry1964general}, Apollo powered descent guidance \cite{klumpp1974apollo}, Zero-Effort-Miss/Zero-Effort-Velocity (ZEM/ZEV) guidance \cite{guo2013applications, simplicio2019guidance}, and real-time trajectory optimization \cite{lu2018propellant, reynolds2020dual, Sagliano2024Six, Malyuta2022Convex}.
However, there is a problem difficult to deal with: as the time-to-go tends to zero, the sensitivity of the generated guidance command trajectories will significantly increase and inevitably tend to infinity;
in the presence of persistent disturbances, this inherent problem will lead to the guidance command dispersion tending to infinity in the terminal time.
In conjunction with the real-time trajectory optimization, the trajectory tracking guidance method \cite{lopez2018robust, shen2010desensitizing} is a widely-used technology route to attenuate disturbances.
The trajectory tracking guidance method can attenuate disturbances by designing the closed-loop tracking control law, but it is hard to achieve the specified landing accuracy due to lacking the direct map between the tracking properties and the trajectory dispersion requirements.
Besides, the MPC guidance method \cite{Lee2017Constrained, bonaccorsi2022dynamic} converts the landing guidance problem into a typical MPC problem and reduce the adverse effects of disturbances by carefully designing three components: terminal control law, terminal set, and cost function.
Nevertheless, it is difficult to design suitable components that meet the desired trajectory dispersion, especially in the case of nonlinear dynamics of the endoatmospheric landing.
By and large, although the above mentioned deterministic guidance methods have robustness to disturbances, they do not directly address the trajectory dispersion control issue.

To achieve the trajectory dispersion control for precision landing, the probabilistic guidance methods have been studied in recent years, including covariance control guidance and robust trajectory optimization.
The covariance control guidance method achieves trajectory dispersion control by steering a linear dynamics system with additive white Gaussian noise from an initial state dispersion to a desired one at a prescribed time.
For example, a feedback control law is designed to constrain the covariance of the terminal state, and the thrust dispersion is controlled within the permissible limits with a high probability \cite{exarchos2019optimal, ridderhof2021minimum}.
A chance constraint is designed to restrict the magnitude of the closed-loop control within a specified probability level, and a convexification strategy is developed to recast the nonlinear covariance control problem as a deterministic convex optimization problem \cite{Benedikter2022, benedikter2023convex}.
In short, the covariance control guidance method enables trajectory dispersion control for the linear dynamics with white Gaussian noise, and exhibits high landing accuracy in the aerodynamic force-free landing problem.
However, since the significant disturbances in the dense atmosphere are difficult to be described by white Gaussian model, it is challenging to directly apply this method to the endoatmospheric landing guidance problem.
Considering more complex disturbances, the robust trajectory optimization can be used to reduce trajectory dispersion by modifying nominal trajectories and guidance parameters.
A robust trajectory optimization procedure based on the polynomial chaos expansion technique is proposed to make the nominal trajectory less sensitive to disturbances \cite{wang2019robust}.
A genetic algorithm is desigend to determine guidance parameters to minimize the impact of initial condition, environment, navigation, and vehicle property uncertainty on flight performance \cite{doi:10.2514/1.A35845}.
Overall, the above robust trajectory optimization methods achieve trajectory dispersion control by solving complex optimization problems offline.
However, this kind of trajectory dispersion control method is usually conservative due to the presence of initial state uncertainty;
in actual flight, the current state is deterministic, and the trajectory dispersion starting from the current state will gradually decrease as the time-to-go reduces.
Therefore, the landing guidance performance can be further improved by online trajectory dispersion prediction and control.

In this note, a novel online trajectory dispersion control method is proposed to achieve precision landing by directly shaping the trajectory dispersions of both states and commands in real time.
Based on a Parameterized Optimal Feedback Guidance Law (POFGL), two key components of the proposed method are designed: online trajectory dispersion prediction and real-time guidance parameter tuning for trajectory dispersion optimization.
First, by formalizing a parameterized probabilistic disturbance model, the closed-loop trajectory dispersion under the POFGL is predicted online.
Compared with the covariance control guidance method, a more accurate trajectory dispersion prediction is achieved by using generalized Polynomial Chaos (gPC) expansion and pseudospectral collocation methods.
Second, to ensure computational efficiency, a gradient descent based real-time guidance parameter tuning law is designed to simultaneously optimize the performance index and meet the landing error dispersion constraint, which significantly reduces the conservativeness of guidance design compared with the robust trajectory optimization method.
Simulation results show that the trajectory dispersion prediction method has the same high accuracy as Monte Carlo method, but the computational resource consumption is much smaller than Monte Carlo method;
the real-time guidance parameter tuning law can improve the optimal performance index and meet the desired landing accuracy requirements.

\section{Problem Formulation}
\label{Sec-2}

In this section, to accurately describe the trajectory dispersion control problem, a nonlinear dynamics model of the reusable rocket is established, and a probabilistic disturbance model is proposed.
At last, the trajectory dispersion control problem is formulated as a stochastic optimal control problem.

\subsection{Nonlinear Dynamics Model with Disturbances}

\begin{figure}[!h]
\centering
\includegraphics[width=0.5\linewidth]{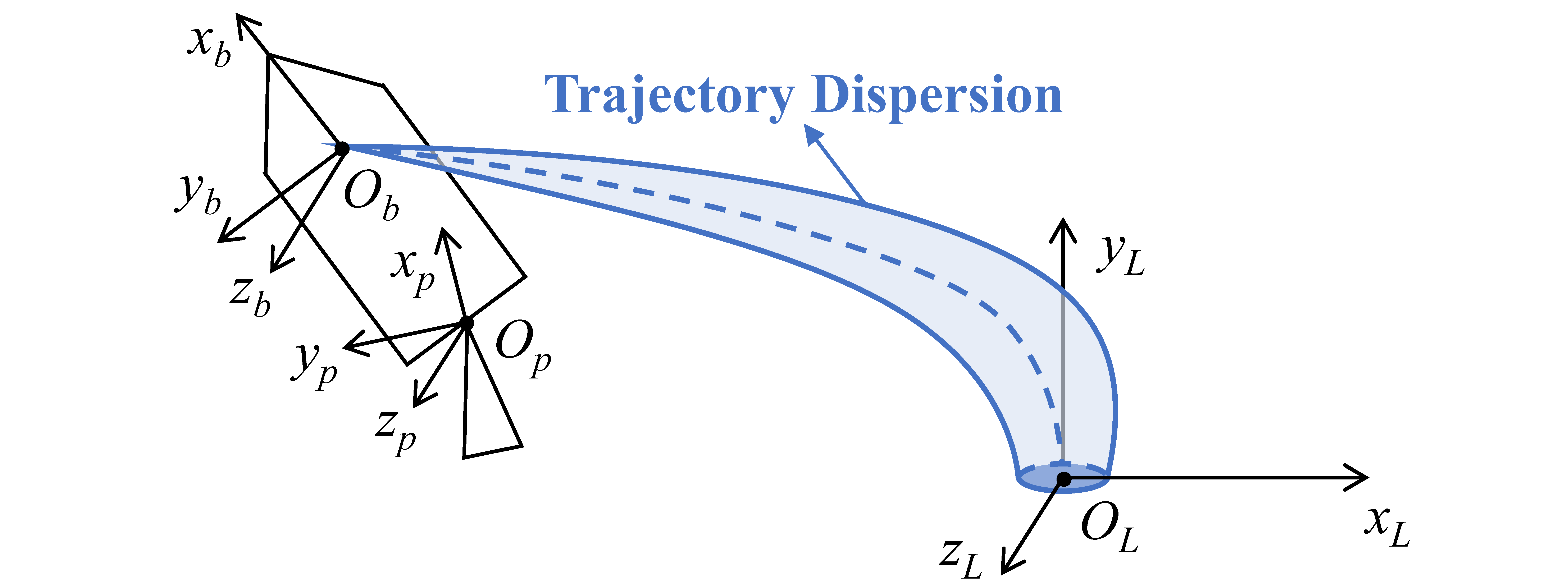}
\caption{Schematic diagram of rocket trajectory dispersion.}
\label{fig-pdg}
\end{figure}

As shown in \cref{fig-pdg}, the rocket's trajectory will exhibit a dispersion with flight time in the presence of disturbances.
To describe the nonlinear dynamics model with disturbances, three coordinate frames are defined as shown in \cref{fig-pdg}.
The inertially-fixed frame \(S_L\) is established with the targeted landing point as its origin, where \(x_L\), \(y_L\) and \(z_L\) axes point to north, up and east.
The body-fixed frame \(S_b\) is established with the rocket's centre of mass as its origin, where \(x_b\), \(y_b\) and \(z_b\) axes point to   forward, upward and rightward.
The thrust-vector-fixed frame \(S_p\) is established with the engine thrust effect point as its origin, where \(x_p\), \(y_p\) and \(z_p\) axes point to forward, upward and rightward.

In the inertially-fixed frame \(S_L\), the 6-DoF dynamics model of the reusable rocket is given \cite{Simplcio2020Reusable}.
The scenario in this paper is assumed that the engine is working in the whole powered descent phase.
Thus, to reduce model complexity, the dynamics model is simplified by neglecting the rotational dynamics and using the trimmed thrust vector angles.
\begin{align}
\label{eq:dyn-0-r}
&\dot{\vect{r}}(t) = \vect{v}(t)
\\
\label{eq:dyn-0-v}
&\dot{\vect{v}}(t)
= \vect{g}(\vect{r})
+\frac{1}{m(t)}[\vect{F}_A(\vect{r}, \vect{v}, \varphi, \psi, \vect{w})
+\vect{F}_{T}(\vect{r}, \vect{v}, \varphi, \psi, u_T, \vect{w})]
\\
\label{eq:dyn-0-m}
&\dot{m}(t) = -\frac{\bar{T} + d_{T}(t)}{V_\mathrm{ex}}u_T(t)
\\
\label{eq:dyn-0-varphi}
&\dot{\varphi}(t) = \omega_\varphi(t)
\\
\label{eq:dyn-0-psi}
&\dot{\psi}(t) = \omega_\psi(t)
\end{align}
where
\(t\in[t_0, t_\mathrm{f}]\) is the time;
\(t_0\) is the given initial time;
\(t_\mathrm{f}\) is the unknown terminal time;
\(\vect{r}(t)\) is the position vector;
\(\vect{v}(t)\) is the velocity vector;
\(m(t)\) is the mass;
\(\varphi(t)\) is the pitch angle command;
\(\psi(t)\) is the yaw angle command;
\(u_T(t)\) is the engine throttling ratio;
\(\omega_\varphi(t)\) is the pitch angular rate;
\(\omega_\psi(t)\) is the yaw angular rate;
\(\vect{g}\) is the gravitational acceleration vector which uses a spherical gravity field model;
\(\vect{F}_A\) is the aerodynamic force vector;
\(\vect{F}_T\) is the engine thrust vector;
\(\bar{T}\) is the constant nominal thrust;
\(V_\mathrm{ex}\) is the constant exhaust velocity;
\(\vect{w}(t)\) is the disturbance vector, includes the engine thrust deviation \(d_{T}(t)\), the attitude angle tracking deviations \(d_{\varphi}(t)\) and \(d_{\psi}(t)\), the aerodynamic coefficient deviations \(d_{Cx}(t)\), \(d_{Cy}(t)\) and \(d_{Cz}(t)\), the atmospheric density deviation \(d_{\rho}(t)\), and the wind disturbance \(\vect{v}_w(t)\).
The disturbance vector \(\vect{w}(t)\) is expressed as
\(
\label{w}
\vect{w}(t) =
\left[
d_{T}(t) \quad d_{\varphi}(t) \quad d_{\psi}(t) \quad
d_{Cx}(t) \quad d_{Cy}(t) \quad d_{Cz}(t) \quad
d_{\rho}(t) \quad \vect{v}_w^\mathrm{T}(t)
\right]^\mathrm{T}
\).

The aerodynamic force vector \(\vect{F}_A\) is
\begin{equation}
\vect{F}_A = \frac{1}{2}[\bar{\rho}(\vect{r}) + d_{\rho}(t)]\norm{\vect{v}_c(t)}^2S_\mathrm{ref}
\vect{R}^\mathrm{T}_{bL}(\varphi_a, \psi_a)
\left[
\bar{C}_x(\alpha, \beta) + d_{Cx}(t)
\quad
\bar{C}_y(\alpha, \beta) + d_{Cy}(t)
\quad
\bar{C}_z(\alpha, \beta) + d_{Cz}(t)
\right]^\mathrm{T}
\end{equation}
where
\(\vect{v}_c(t) = \vect{v}(t) - \vect{v}_w(t)\) is the velocity relative to the atmosphere;
\(\varphi_a = \varphi + d_{\varphi}\) is the true pitch angle;
\(\psi_a = \psi + d_{\psi}\) is the true yaw angle;
\(S_\mathrm{ref}\) is the reference area;
\(\vect{R}_{bL}\) is the rotation matrix from the frame \(S_L\) to the frame \(S_b\);
\(\bar{C}_x\), \(\bar{C}_y\) and \(\bar{C}_z\) are the nominal aerodynamic coefficients;
\(\alpha(t)\) and \(\beta(t)\) are the attack angle and the sideslip angle, respectively.

The engine thrust vector \(\vect{F}_T\) is
\begin{equation}
\vect{F}_{T} = [\bar{T} + d_{T}(t)]\vect{R}^\mathrm{T}_{bL}(\varphi_a, \psi_a)\vect{R}^\mathrm{T}_{pb}(\delta_\varphi, \delta_\psi)
\left[
u_T(t)
\quad
0
\quad
0
\right]^\mathrm{T}
\end{equation}
where
\(\vect{R}_{pb}\) is the rotation matrix from the frame \(S_b\) to the frame \(S_p\);
\(\delta_\varphi\) and \(\delta_\psi\) are the trimmed thrust vector angles in the pitch direction and yaw direction, respectively.

To reduce the model's complexity and nonlinearity, the rotational dynamics is neglected, and the trimmed thrust vector angles \(\delta_\varphi\) and \(\delta_\psi\) are used to balance aerodynamic torques and thrust torques, denoted as
\begin{equation}
\label{delta-varphi}
\delta_\varphi \approx {M_{Az}(\vect{r}, \vect{v}, \varphi, \psi)}/[{\bar{T}r_Tu_T(t)}]
,\quad
%\label{delta-psi}
\delta_\psi \approx {M_{Ay}(\vect{r}, \vect{v}, \varphi, \psi)}/[{\bar{T}r_Tu_T(t)}]
\end{equation}
where \(r_T\) is the distance between the centre of mass and the point of engine thrust effect;
\(M_{Ay}\) and \(M_{Az}\) are the aerodynamic torques around the \(y_b\) and \(z_b\) axes.

Define the state vector as
\(
\vect{x}(t) =
\left[
\vect{r}^\mathrm{T}(t) \quad \vect{v}^\mathrm{T}(t) \quad m(t) \quad \varphi(t) \quad \psi(t)
\right]^\mathrm{T}
\).
Define the guidance command vector as
\(
\vect{u}(t) =
\left[
u_T(t) \quad \omega_\varphi(t) \quad \omega_\psi(t)
\right]^\mathrm{T}
\).
Define the total flight time as \(a = t_\mathrm{f} - t_0\) and normalize the time as
\(
\tau = ({t - t_0})/{a} \in [0, 1]
\),
where \(\tau\) is called the normalized time.
Denoting the differential symbol as the derivative of the normalized time \(\tau\),
Eqs. (\ref{eq:dyn-0-r}--\ref{eq:dyn-0-psi}) can be written in the compact form as
\begin{equation}
\label{eq:dyn-x}
\dot{\vect{x}}(\tau) = a\tilde{\vect{f}}[\tau, \vect{x}(\tau), \vect{u}(\tau), \vect{w}(\tau)]
\triangleq \vect{f}[\tau, \vect{x}(\tau), \vect{u}(\tau), a, \vect{w}(\tau)]
\end{equation}

In an actual flight mission, the state \(\vect{x}(t)\) is available by the navigation system, and the disturbance \(\vect{w}(t)\) is certain but unknown;
in different flights, the trajectories will exhibit uncertain dispersion under the effect of disturbances.

\subsection{Probabilistic Disturbance Model}

\begin{figure*}[!h]
        \centering
    \begin{subfigure}[h]{0.45\linewidth}
        \centering
        \includegraphics[width=\textwidth]{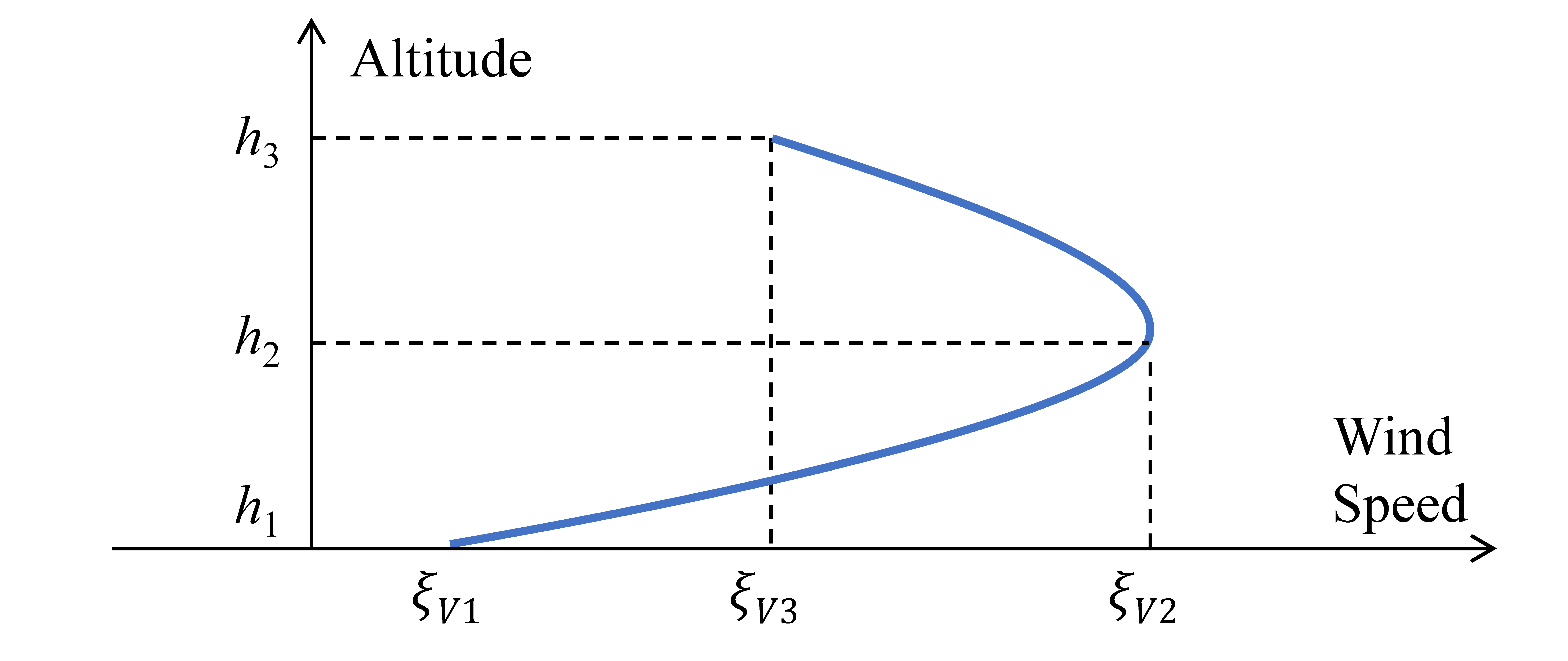}
        \caption{Single wind field model.}
    \end{subfigure}
	 \begin{subfigure}[h]{0.45\linewidth}
        \centering
        \includegraphics[width=\textwidth]{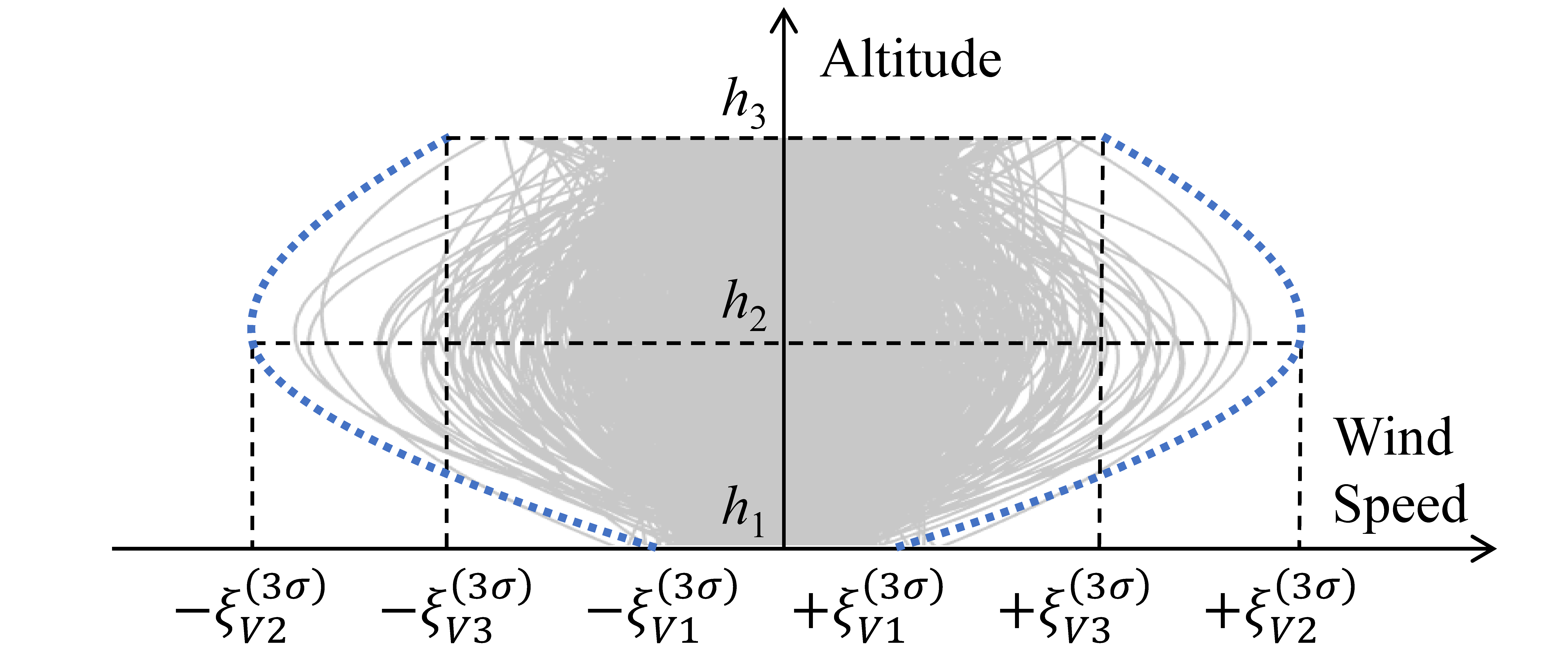}
        \caption{Wind field dispersion model.}
    \end{subfigure}
    \caption{Schematic diagram of wind field model using polynomial fitting (\(M = 3\)).}
    \label{fig-wind}
\end{figure*}

In this subsection, to describe the unknown and unpredictable disturbances of the endoatmospheric landing guidance, a parameterized probabilistic disturbance model is formulated.
Borrowing from the disturbance setting in Monte Carlo method \cite{doi:10.2514/1.48813}, the disturbance vector \(\vect{w}(t)\) can be modeled as a function of random variables as
\begin{equation}
\label{w(t)}
\vect{w}(t) =
\left[
\xi_{T}\bar{T} \quad \xi_{\varphi} \quad \xi_{\psi} \quad
\xi_{Cx}\bar{C}_x(\alpha, \beta) \quad \xi_{Cy}\bar{C}_y(\alpha, \beta) \quad \xi_{Cz}\bar{C}_z(\alpha, \beta)  
\quad
\xi_{\rho}\bar{\rho}(\vect{r}) \quad \vect{p}_{vw}^{T}(h)
\right]^\mathrm{T}
\end{equation}
where
\(\xi_{T}\) is the random variable related to the thrust deviation;
\(\xi_{\varphi}\) and \(\xi_{\psi}\) are the random variables related to the attitude angle tracking deviations;
\(\xi_{Cx}\), \(\xi_{Cy}\), and \(\xi_{Cz}\) are the random variables related to the aerodynamic coefficient deviations;
\(\xi_{\rho}\) is the random variable related to the atmospheric density deviation;
\(h = \norm{\vect{R}_E + \vect{r}(t)} - R_\mathrm{earth}\) is the flight altitude;
\(\vect{R}_E\) is the position vector from the center of the Earth to the origin of the inertially-fixed frame \(S_L\);
\(R_\mathrm{earth}\) is the average radius of the Earth;
\(\vect{p}_{vw}\) is the random wind field model using polynomial fitting as shown in \cref{fig-wind}, denoted as
\begin{equation}
\vect{p}_{vw}(h) =
\left[
\sum_{j = 1}^{M}{\left(\xi_{Vj}\prod_{i = 1, i \neq j}^M{\frac{h-h_i}{h_j - h_i}}\right)}\cos{\xi_{A}}
\quad 0 \quad
\sum_{j = 1}^{M}{\left(\xi_{Vj}\prod_{i = 1, i \neq j}^M{\frac{h-h_i}{h_j - h_i}}\right)}\sin{\xi_{A}}
\right]
\end{equation}
where
\(\xi_{V1}\), \(\xi_{V2}\), \(\cdots\) ,\(\xi_{VM}\) are the random variables related to the wind velocity at the altitudes \(h_{1}\), \(h_{2}\), \(\cdots\) ,\(h_{M}\);
\(\xi_{A}\) is the random variable related to the direction of wind.

Define the random vector as
\(
\vect{\xi}_{w}
=
[\xi_{T} \quad \xi_{\varphi} \quad \xi_{\psi} \quad
\xi_{Cx} \quad \xi_{Cy} \quad \xi_{Cz} \quad
\xi_{\rho} \quad
\xi_{V1} \quad \cdots \quad \xi_{VM} \quad
\xi_{A}
]^\mathrm{T}
\),
where \(\vect{\xi}_{w} \in \mathbb{R}^{n_{\xi w}}\) is a \(n_{\xi w}\)-dimensional independent random vector with a stationary probability density function denoted as
\begin{equation}
\label{pxiw}
p_{\xi w}(\vect{\xi}_{w}) =
p_{\xi w}(\xi_{w1},\cdots,\xi_{wn_{\xi w}}) = 
\prod_{i = 1}^{n_{\xi w}}{p_{\xi wi}(\xi_{wi})}
\end{equation}
where \(p_{\xi wi}\) is the probability density function of the one-dimensional random variable \(\xi_{wi}\).

Different from the additive Gaussian white noise model in the covariance control guidance, the probabilistic disturbance model as Eq. \eqref{w(t)} parameterizes the disturbances as the functions of random variables, which is consistent with the Monte Carlo design: in different flights, different random variables are selected according to the probability density function as Eq. \eqref{pxiw}, resulting in the trajectory dispersion of the rocket.

\subsection{Trajectory Dispersion Control Problem}

Based on the probabilistic disturbance model, the trajectory dispersion control problem can be modeled as a stochastic optimal control problem as follows.
\begin{problem}
\label{Sto-Endo-PDG}
\normalfont
Trajectory Dispersion Control Problem
\begin{alignat}{2}
\label{problem-sto-Js}
&\mathop{\mathrm{minimize}}
\limits_{\vect{\pi}_u[\vect{x}(\tau)],\,\pi_a[\vect{x}(\tau)]}\quad
&&
J_s = \mathrm{E}[-m(1)]
\\
\label{problem-sto-dyn}
&\,\,\,\,\,\,\,\,\,\text{subject to}\quad
&&
\dot{\vect{x}}(\tau) = \vect{f}[\tau, \vect{x}(\tau), \vect{u}(\tau), a, \vect{\xi}_w]
\\
&
&&
\label{problem-sto-x(0)=x0}
\vect{x}(\tau_\mathrm{c}) = \vect{x}_\mathrm{c}
\\
&
&&
\label{problem-sto-terminal-constraint}
\mathrm{Pr}[\vect{C}\vect{x}(1) \in \mathbb{C}_\mathrm{lim}] \geq P_c
\\
&
&&
\label{problem-sto-control-constraint}
\mathrm{Pr}[\vect{u}(\tau) \in \mathbb{U}_\mathrm{lim}] \geq P_u
\\
&
&&
\label{problem-sto-policy}
\vect{u}(\tau) = \vect{\pi}_u[\vect{x}(\tau)],\quad
a = \pi_a[\vect{x}(\tau)]
\end{alignat}
where \(\vect{\pi}_u\) and \(\pi_a\) are the guidance laws to be determined corresponding to the guidance commands and total flight time;
\(\mathrm{E}(\cdot)\) denotes the mean vector;
\(\tau_\mathrm{c}\) is the current normalized time;
\(\vect{x}_\mathrm{c}\) is the current state;
\(\mathrm{Pr}(\cdot)\) denotes the probability of an event;
\(\vect{C}\) is the terminal constraint matrix satisfying
\(\vect{C}\vect{x}(1) =
\left[
\vect{r}(1) \quad
\vect{v}(1) \quad
\varphi(1)\quad
\psi(1)
\right]^\mathrm{T}\);
\(\mathbb{C}_\mathrm{lim} = \{\vect{c} \in \mathbb{R}^8 \mid \vect{c}_\mathrm{min} \leq \vect{c} \leq \vect{c}_\mathrm{max}\}\);
\(P_c\) is the required probability of the terminal constraint;
\(\mathbb{U}_\mathrm{lim} = \{\vect{u} \in \mathbb{R}^3 \mid \vect{u}_\mathrm{min} \leq \vect{u} \leq \vect{u}_\mathrm{max}\}\);
\(P_u\) is the required probability of the control amplitude constraint.
\end{problem}

As shown in \cref{Sto-Endo-PDG}, the object of trajectory dispersion control is optimizing the mean value of the performance index and constraining the terminal state dispersion and the guidance command dispersion within the given probability ranges.
The probabilistic trajectory dispersion of the state and the guidance command can be characterized using mean value and variance value.
Generally, \cref{Sto-Endo-PDG} is hard to solve for two main reasons:
first, solving for the feedback guidance laws \(\vect{\pi}_u\) and \(\pi_a\) in a high-dimensional continuousspace system has ``the curse of dimensionality'';
second, the presence of nonlinear dynamics significantly complicates the process of obtaining an online solution.

\begin{figure}[!t]
\centering
\includegraphics[width=0.7\linewidth]{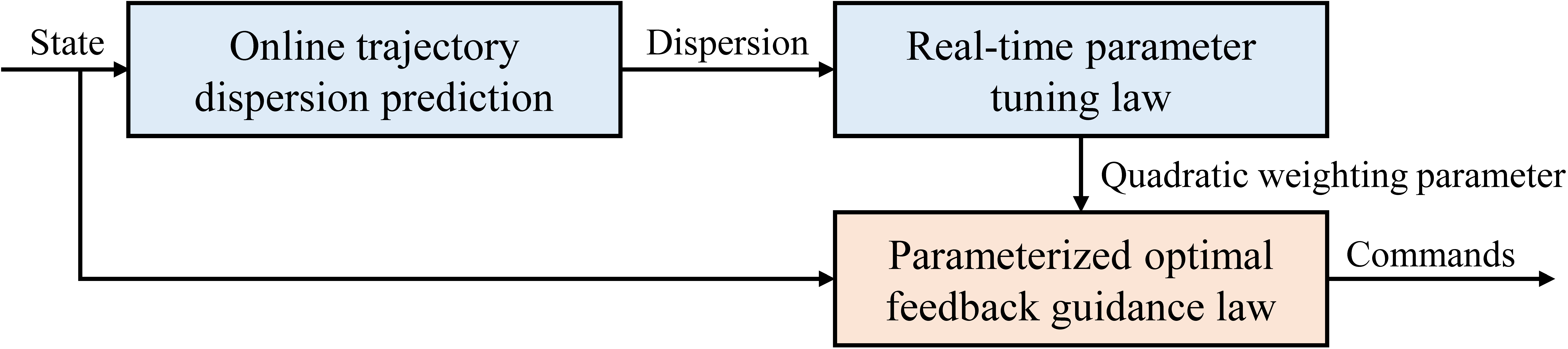}
\caption{Framework of trajectory dispersion control.}
\label{fig-framework}
\end{figure}

To address these issues, a trajectory dispersion control framework is proposed as shown in \cref{fig-framework}.
The main part of the framework is a Parameterized Optimal Feedback Guidance Law (POFGL), which can regulate the trajectory dispersion by tuning parameterized time-varying weights.
Based on the POFGL, this framework consists of two key designs.
First, in \cref{Sec-4}, the trajectory dispersion is predicted online by approximating the stochastic closed-loop dynamics system as a higher-dimensional deterministic dynamics system using gPC expansion and pseudospectral collocation methods.
Subsequently, \cref{Sto-Param-Opt} can be transformed into a deterministic parameter optimization problem, and, in \cref{Sec-5}, a real-time guidance parameter tuning law based on gradient descent method is designed to optimize the trajectory dispersion by solving the deterministic parameter optimization problem.
In this framework, the POFGL can be replaced with other parameterized guidance laws, and the trajectory dispersion control can be still achieved through the two aforementioned designs.

\section{Parameterized Optimal Feedback Guidance Law}
\label{Sec-3}

As the main part of the trajectory dispersion control framework, a POFGL is designed in an affine form as
\begin{align}
\label{simguid_law_u}
&\vect{u}(\tau) = \vect{u}_\mathrm{r}(\tau, \vect{x}_0) + \vect{K}_u(\tau, \vect{x}_0, \vect{\theta})[\vect{x}(\tau) - \vect{x}_\mathrm{r}(\tau, \vect{x}_0)]
\\
&\label{simguid_law_a}
a(\tau) = a_\mathrm{r}(\vect{x}_0) + \vect{K}_a(\tau, \vect{x}_0, \vect{\theta})[\vect{x}(\tau) - \vect{x}_\mathrm{r}(\tau, \vect{x}_0)]
\end{align}
where \(\vect{x}_\mathrm{r}(\tau, \vect{x}_0)\), \(\vect{u}_\mathrm{r}(\tau, \vect{x}_0)\), and \(a_\mathrm{r}(\vect{x}_0)\) are nominal trajectories calculated by solving \cref{Nom-Endo-PDG} below;
\(\vect{K}_u(\tau, \vect{x}_0, \vect{\theta})\) and \(\vect{K}_a(\tau, \vect{x}_0, \vect{\theta})\) are feedback matrices calculated by solving \cref{Par-Endo-PDG} below;
\(\vect{\theta}\) is the quadratic weighting parameter.

\begin{problem}
\label{Nom-Endo-PDG}
\normalfont
Optimal Control Problem for Determining Nominal Trajectory of POFGL
\begin{alignat}{2}
\label{problem-nom-J0}
&\mathop{\mathrm{minimize}}
\limits_{\vect{x}(\tau),\,\vect{u}(\tau),\,a}\quad
&&
J_0 =
- m(1)
+ [\vect{C}\vect{x}(1) - \vect{c}_\mathrm{f}]^\mathrm{T}\vect{R}_1[\vect{C}\vect{x}(1) - \vect{c}_\mathrm{f}]
+ a\int_{0}^{1}{[\vect{u}(\tau) - \vect{u}_\mathrm{m}]^\mathrm{T}\vect{R}_0[\vect{u}(\tau) - \vect{u}_\mathrm{m}]\mathrm{d}\tau}
\\
&\,\,\,\text{subject to}\quad
&&
\label{problem-nom-dyn}
\dot{\vect{x}}(\tau) = \bar{\vect{f}}[\tau, \vect{x}(\tau), \vect{u}(\tau), a]
\\
&
&&
\label{problem-nom-x(0)=x0}
\vect{x}(0) = \vect{x}_0
\end{alignat}
where \(\bar{\vect{f}}[\tau, \vect{x}(\tau), \vect{u}(\tau), a]\) is the nominal dynamics equation with \(\vect{\xi}_{w} = \vect{0}\);
\(\vect{u}_\mathrm{m}\) is the median value of the allowed guidance command amplitude, denoted as
\(
\vect{u}_\mathrm{m} = (\vect{u}_\mathrm{min} + \vect{u}_\mathrm{max})/2
\);
\(\vect{u}_\mathrm{min}\) is the minimum guidance command;
\(\vect{u}_\mathrm{max}\) is the maximum guidance command;
\(\vect{c}_\mathrm{f} = [\vect{0} \quad \vect{0} \quad \pi/2 \quad 0]^\mathrm{T}\);
\(\vect{R}_0\) and \(\vect{R}_1\) are the positive definite diagonal matrices.
\end{problem}

\begin{remark}
\textnormal{
The quadratic terms in Eq. \eqref{problem-nom-J0} are soft constraints  corresponding to the control amplitude hard constraint and the terminal hard constraint.
There are three reasons for using the soft constraints:
the modeling strategy using the soft constraints can simplify the solution of the problem;
the control amplitude soft constraint ensures the smoothness of the control, thereby significantly reducing angular accelerations \(\dot{\omega}_\varphi(t)\) and \(\dot{\omega}_\psi(t)\);
the terminal soft constraint can circumvent the infinity of guidance command sensitivity.
}
\end{remark}

\begin{problem}
\label{Par-Endo-PDG}
\normalfont
Optimal Control Problem for Determining Feedback Matrices of POFGL
\begin{alignat}{2}
\label{problem-par-J0}
&\mathop{\mathrm{minimize}}
\limits_{\vect{\pi}_u[\vect{x}(\tau)],\,\pi_a[\vect{x}(\tau)]}\quad
&&
J = J_0 + J_1
\\
&\,\,\,\,\,\,\,\,\,\text{subject to}\quad
&&
\text{Eqs. \eqref{problem-nom-dyn} and \eqref{problem-nom-x(0)=x0}}
\nonumber
\\
&
&&
\label{problem-par-policy}
\vect{u}(\tau) = \vect{\pi}_u[\vect{x}(\tau)],\quad
a = \pi_a[\vect{x}(\tau)]
\\
&
&&
\label{problem-par-xrurar}
\text{\(\vect{x}_\mathrm{r}(\tau)\), \(\vect{u}_\mathrm{r}(\tau)\), and \(a_\mathrm{r}\) satisfy the solution of \cref{Nom-Endo-PDG} with \(\vect{x}_\mathrm{r}(0) = \vect{x}_0\)}
\end{alignat}
where
\begin{equation}
\label{J_1}
J_1
= 
\updelta\vect{x}^\mathrm{T}(1)\vect{R}_\mathrm{f}(\vect{\theta})\updelta\vect{x}(1)
+
\int_{0}^{1}
{
\begin{bmatrix}
\updelta\vect{x}(\tau)
\\
\updelta\vect{u}(\tau)
\\
\mathrm{d}a
\end{bmatrix}^\mathrm{T}
\begin{bmatrix}
\vect{R}_x(\tau, \vect{\theta}) & \vect{0} & \vect{0}
\\
\vect{0} & \vect{R}_u(\tau, \vect{\theta}) & \vect{0}
\\
\vect{0} & \vect{0} & R_a(\tau, \vect{\theta})
\end{bmatrix}
\begin{bmatrix}
\updelta\vect{x}(\tau)
\\
\updelta\vect{u}(\tau)
\\
\mathrm{d}a
\end{bmatrix}
\mathrm{d}\tau
}
\end{equation}
where
\(\updelta\vect{x}(\tau) = \vect{x}(\tau) - \vect{x}_\mathrm{r}(\tau)\);
\(\updelta\vect{u}(\tau) = \vect{u}(\tau) - \vect{u}_\mathrm{r}(\tau)\);
\(\mathrm{d}a = a - a_\mathrm{r}\);
\(\vect{R}_\mathrm{f}(\vect{\theta})\) is the positive definite matrix;
\(\vect{R}_x(\tau, \vect{\theta})\), \(\vect{R}_u(\tau, \vect{\theta})\) and \(R_a(\tau, \vect{\theta})\) are the positive definite matrices.

The weights matrix \(\vect{R}_\mathrm{f}(\vect{\theta})\) is parameterized as \(\vect{R}_\mathrm{f}(\vect{\theta}) \triangleq \vect{R}_{\theta\mathrm{f}}^\mathrm{T}\vect{R}_{\theta\mathrm{f}}\), and the time-varying weights matrices \(\vect{R}_x(\tau, \vect{\theta})\), \(\vect{R}_u(\tau, \vect{\theta})\) and \(R_a(\tau, \vect{\theta})\) are parameterized  as
\begin{equation}
\vect{R}_x(\tau, \vect{\theta})
=
\sum_{j = 1}^{M}
{\left(
\vect{R}_{xj}
\prod_{i = 1, i \neq j}^{M}
{\frac{\tau - \tau_i}{\tau_j - \tau_i}}
\right)}
,\quad
\vect{R}_u(\tau, \vect{\theta})
=
\sum_{j = 1}^{M}
{\left(
\vect{R}_{uj}
\prod_{i = 1, i \neq j}^{M}
{\frac{\tau - \tau_i}{\tau_j - \tau_i}}
\right)}
,\quad
R_a(\tau, \vect{\theta})
=
\sum_{j = 1}^{M}
{\left(
R_{aj}
\prod_{i = 1, i \neq j}^{M}
{\frac{\tau - \tau_i}{\tau_j - \tau_i}}
\right)}
\end{equation}
where \(\vect{R}_{xj} \triangleq \vect{R}_{\theta xj}^\mathrm{T}\vect{R}_{\theta xj}\), \(\vect{R}_{uj} \triangleq \vect{R}_{\theta uj}^\mathrm{T}\vect{R}_{\theta uj}\), and \({R}_{aj} \triangleq {R}_{\theta aj}^2\) are positive semidefinite matrices.
The quadratic weighting parameter vector \(\vect{\theta}\) is defined as
\begin{equation}
\vect{\theta} \triangleq
[
\mathrm{vect}(\vect{R}_{\theta\mathrm{f}})
\quad
\mathrm{vect}(\vect{R}_{\theta x1})
\quad
\cdots
\quad
\mathrm{vect}(\vect{R}_{\theta xM})
\quad
\mathrm{vect}(\vect{R}_{\theta u1})
\quad
\cdots
\quad
\mathrm{vect}(\vect{R}_{\theta uM})
\quad
{R}_{\theta a1}
\quad
\cdots
\quad
{R}_{\theta aM}
]^\mathrm{T}
\end{equation}
\end{problem}

\begin{remark}
\textnormal{
The particularity of \cref{Par-Endo-PDG} is the addition of the parameterized time-varying quadratic performance index \(J_1\).
Intuitively, the added quadratic performance index \(J_1\) is a damping term that encourages \(\vect{x}(\tau)\), \(\vect{u}(\tau)\), and \(a\) not to be very far from \(\vect{x}_\mathrm{r}(\tau)\), \(\vect{u}_\mathrm{r}(\tau)\), and \(a_\mathrm{r}\).
By tuning the parameterized time-varying weights, trajectory dispersion and landing accuracy can be regulated \cite{Paper-Optimal}.
The feedback guidance law as Eqs. \eqref{simguid_law_u} and \eqref{simguid_law_a} is parameterized through time-varying weights \(R_a(\tau, \vect{\theta})\), \(\vect{R}_x(\tau, \vect{\theta})\), \(\vect{R}_u(\tau, \vect{\theta})\) and \(\vect{R}_\mathrm{f}(\vect{\theta})\), instead of directly varying feedback coefficients \(\vect{K}_u(\tau)\) and \(\vect{K}_a(\tau)\).
This form of parameterization provides two key advantages:
first, in the absence of disturbances, \(J_1\) does not affect the optimality of the trajectory, and the guidance law represents a form of neighboring optimal guidance law that ensures terminal constraint;
second, parametrizing time-varying weights is helpful to focus on most relevant parameters and to restrict the search space of the guidance law.
}
\end{remark}

To solve \cref{Nom-Endo-PDG} and \cref{Par-Endo-PDG}, a Pseudospectral Differential Dynamic Programming (PDDP) method \cite{Paper-Optimal} is developed.
This method can simultaneously calculate the nominal optimal trajectory and the feedback coefficients in Eqs. \eqref{simguid_law_u} and \eqref{simguid_law_a} by iteratively solving the second-order expansion of the Hamilton-Jacobi-Bellman (HJB) equation of \cref{Nom-Endo-PDG} and \cref{Par-Endo-PDG}.
In the PDDP method, \(\vect{x}_\mathrm{r}(\tau)\), \(\vect{u}_\mathrm{r}(\tau)\), \(a_\mathrm{r}\), \(\vect{K}_u(\tau, \vect{\theta})\) and \(\vect{K}_a(\tau, \vect{\theta})\) are calculated at the initial time with \(\vect{x}(0) = \vect{x}_\mathrm{r}(0) = \vect{x}_0\),
and the feedback guidance law as Eqs. \eqref{simguid_law_u} and \eqref{simguid_law_a} is implemented at the subsequent time with \(\vect{x}_\mathrm{r}(\tau)\), \(\vect{u}_\mathrm{r}(\tau)\), and \(a_\mathrm{r}\) fixed.
More importantly, the numerical computations of the PDDP method are performed within a pseudospectral setting, so that \(\vect{x}_\mathrm{r}(\tau)\), \(\vect{u}_\mathrm{r}(\tau)\), \(\vect{K}_u(\tau, \vect{\theta})\), and \(\vect{K}_a(\tau, \vect{\theta})\) are represented as the analytical orthogonal polynomial functions of the normalized time.

Based on the POFGL, \cref{Sto-Endo-PDG} can be transformed into the following parameter optimization problem.

\begin{problem}
\label{Sto-Param-Opt}
\normalfont
Trajectory Dispersion Control Problem Based on POFGL
\begin{alignat}{2}
&\mathop{\mathrm{minimize}}
\limits_{\vect{\theta} \in \vect{\Theta}}\quad
&&
\text{Eq. \eqref{problem-sto-Js}}
\nonumber
\\
&\,\,\text{subject to}\quad
&&
\text{Eqs. \eqref{problem-sto-dyn}--\eqref{problem-sto-control-constraint} \eqref{simguid_law_u} and \eqref{simguid_law_a}}
\nonumber
\end{alignat}
where \(\vect{\Theta}\) is the set of allowable values of \(\vect{\theta}\).
\end{problem}

\cref{Sto-Param-Opt} is a parameter optimization problem with stochastic dynamics and probabilistic constraints.
The POFGL and reusable rocket dynamics constitute a closed-loop dynamics system.
Varying the quadratic weighting parameter of the POFGL can change the closed-loop trajectories in the presence of disturbances, thereby affecting the trajectory dispersion.
To solve \cref{Sto-Param-Opt}, the trajectory dispersion of the closed-loop stochastic dynamics system is predicted in \cref{Sec-4}, and the trajectory dispersion control is achieved through an online parameter tuning law in \cref{Sec-5}.

\section{POFGL Based Online Trajectory Dispersion Prediction}
\label{Sec-4}

The key for achieving trajectory dispersion control is the online trajectory dispersion prediction.
In this section, taking the current state as a starting point, the trajectory dispersion is predicted online by approximating the stochastic closed-loop dynamics system as a higher-dimensional deterministic dynamics system.

Substituting Eqs. \eqref{simguid_law_u} and \eqref{simguid_law_a} into Eq. \eqref{eq:dyn-x}, the closed-loop stochastic dynamics system is obtained as
\begin{equation}
\label{sto_eq:dyn-x}
\dot{\vect{x}}(\tau) = \vect{f}\{\tau, \vect{x}(\tau), \vect{u}[\tau, \vect{x}(\tau), \vect{x}_0, \vect{\theta}], a[\tau, \vect{x}(\tau), \vect{x}_0, \vect{\theta}], \vect{\xi}_w\}
\triangleq
\vect{F}[\tau, \vect{x}(\tau), \vect{\xi}_w, \vect{\theta}]
\end{equation}

By applying the gPC expansion to finite order of \(\vect{x}(\tau)\), the closed-loop state \(\vect{x}(\tau)\) can be approximated as
\begin{equation}
\label{pce_x}
\vect{x}(\tau, \vect{\xi}_w, \vect{\theta})
=
\sum_{k = 0}^{\infty}{\vect{x}_k^c(\tau, \vect{\theta})\phi_k(\vect{\xi}_w)}
\approx
\sum_{k = 0}^{P}{\vect{x}_k^c(\tau, \vect{\theta})\phi_k(\vect{\xi}_w)}
\end{equation}
where
\(\phi_k(\vect{\xi}_w)\) is the gPC basis of degree \(k\) in terms of the random variable \(\vect{\xi}_w\);
\(\vect{x}_k^c(\tau, \vect{\theta})\) is the expansion coefficient of degree \(k\);
\(P\) is the number of truncated terms.
Substituting Eq. \eqref{pce_x} into Eq. \eqref{sto_eq:dyn-x} yields
\begin{equation}
\label{dot_nonlinear_pce_x}
\sum_{k = 0}^{P}{\dot{\vect{x}}_k^c(\tau, \vect{\theta})\phi_k(\vect{\xi}_w)} =
\vect{F}
\left[\tau, \sum_{k = 0}^{P}{\vect{x}_k^c(\tau, \vect{\theta})\phi_k(\vect{\xi}_w)}, \vect{\xi}_w, \vect{\theta}\right]
\end{equation}

The expansion coefficient \(\vect{x}_k^c(\tau, \vect{\theta})\) can usually be calculated using Galerkin projection and numerical integration based on Eq. \eqref{dot_nonlinear_pce_x}.
To improve the efficiency of online solving, an approximate calculation method for the expansion coefficients is presented by solving a higher-dimensional deterministic linear dynamics system in the pseudospectral setting, and the expansion coefficient \(\vect{x}_k^c(\tau, \vect{\theta})\) can be obtained as an analytical function of \(\vect{\theta}\).

First, Eq. \eqref{dot_nonlinear_pce_x} is approximated to a higher-dimensional deterministic dynamics system.
Take expansion of Eq. \eqref{dot_nonlinear_pce_x} around \(\vect{x}_\mathrm{r}(\tau, \vect{x}_0)\), \(\vect{u}_\mathrm{r}(\tau, \vect{x}_0)\), and \(a_\mathrm{r}(\vect{x}_0)\) to first-order as
\begin{equation}
\label{sto_lin_dyn}
\dot{\vect{x}}(\tau, \vect{\xi}_w, \vect{\theta})
= \vect{A}(\tau, \vect{\xi}_w, \vect{\theta})\vect{x}(\tau, \vect{\xi}_w, \vect{\theta})
+ \vect{b}(\tau, \vect{\xi}_w, \vect{\theta})
\end{equation}
where
\(\vect{A}(\tau, \vect{\xi}_w, \vect{\theta})\) and \(\vect{b}(\tau, \vect{\xi}_w, \vect{\theta})\) can be expressed as the analytical forms of \(\tau\), \(\vect{\theta}\) and \(\vect{\xi}_w\), since \(\vect{x}_\mathrm{r}(\tau, \vect{x}_0)\), \(\vect{u}_\mathrm{r}(\tau, \vect{x}_0)\), \(\vect{K}_u(\tau, \vect{x}_0, \vect{\theta})\), and \(\vect{K}_a(\tau, \vect{x}_0, \vect{\theta})\) have been represented as the analytical orthogonal polynomial functions of \(\tau\) and \(\vect{\theta}\).

Using gPC expansion, \(\vect{A}(\tau, \vect{\xi}_w, \vect{\theta})\) and \(\vect{b}(\tau, \vect{\xi}_w, \vect{\theta})\) can be approximated as
\begin{equation}
\label{pce_A_pce_b}
\vect{A}(\tau, \vect{\xi}_w, \vect{\theta})
\approx
\sum_{k = 0}^{P}{\vect{A}_k^c(\tau, \vect{\theta})\phi_k(\vect{\xi}_w)}
,\quad
\vect{b}(\tau, \vect{\xi}_w, \vect{\theta})
\approx
\sum_{k = 0}^{P}{\vect{b}_k^c(\tau, \vect{\theta})\phi_k(\vect{\xi}_w)}
\end{equation}
where
\(\vect{A}_k^c(\tau, \vect{\theta})\) and \(\vect{b}_k^c(\tau, \vect{\theta})\) are the expansion coefficients of degree \(k\), and can be obtained via Galerkin projection onto \(\phi_k(\vect{\xi}_w)\) as
\begin{equation}
\label{pce_Abkc}
\vect{A}_k^c(\tau, \vect{\theta})
=
\frac{\langle\vect{A}(\tau, \vect{\xi}_w, \vect{\theta}),\, \phi_k(\vect{\xi}_w)\rangle}
{\langle\phi_k(\vect{\xi}_w),\, \phi_k(\vect{\xi}_w)\rangle}
,\quad
\vect{b}_k^c(\tau, \vect{\theta})
=
\frac{\langle\vect{b}(\tau, \vect{\xi}_w, \vect{\theta}),\, \phi_k(\vect{\xi}_w)\rangle}
{\langle\phi_k(\vect{\xi}_w),\, \phi_k(\vect{\xi}_w)\rangle}
\end{equation}

In Eq. \eqref{pce_Abkc}, \(\langle\phi_k(\vect{\xi}_w),\, \phi_k(\vect{\xi}_w)\rangle\) can be calculated offline, and \(\langle\vect{A}(\tau, \vect{\xi}_w, \vect{\theta}),\, \phi_k(\vect{\xi}_w)\rangle\) and \(\langle\vect{b}(\tau, \vect{\xi}_w, \vect{\theta}),\, \phi_k(\vect{\xi}_w)\rangle\) can be calculated online as the analytical forms of \(\vect{\theta}\) using analytical or Gaussian integration methods.

Substituting Eqs. \eqref{pce_x} and \eqref{pce_A_pce_b} into Eq. \eqref{sto_lin_dyn} yields
\begin{equation}
\label{pcd_lin_dyn_0}
\sum_{k = 0}^{P}{\dot{\vect{x}}_k^c(\tau, \vect{\theta})\phi_k(\vect{\xi}_w)}
=
\left[
\sum_{k = 0}^{P}{\vect{A}_k^c(\tau, \vect{\theta})\phi_k(\vect{\xi}_w)}
\right]
\left[
\sum_{k = 0}^{P}{\vect{x}_k^c(\tau, \vect{\theta})\phi_k(\vect{\xi}_w)}
\right]
+
\sum_{k = 0}^{P}{\vect{b}_k^c(\tau, \vect{\theta})\phi_k(\vect{\xi}_w)}
\end{equation}
Taking Galerkin projection of Eq. \eqref{pcd_lin_dyn_0} on \(\phi_k(\vect{\xi}_w)\) yields
\begin{equation}
\label{pcd_lin_dyn_1}
\dot{\vect{x}}_k^c(\tau, \vect{\theta})
=
\sum_{i = 0}^{P}
{
\sum_{j = 0}^{P}{\vect{A}_i^c(\tau, \vect{\theta})\vect{x}_j^c(\tau, \vect{\theta})
\frac{\langle\phi_i(\vect{\xi}_w)\phi_j(\vect{\xi}_w),\, \phi_k(\vect{\xi}_w)\rangle}{\langle\phi_k(\vect{\xi}_w),\, \phi_k(\vect{\xi}_w)\rangle}
}
}
+
\vect{b}_k^c(\tau, \vect{\theta})
\end{equation}

Let
\(
\vect{x}_G(\tau, \vect{\theta})
=
[\vect{x}_0^{cT}(\tau, \vect{\theta}) \quad \vect{x}_1^{cT}(\tau, \vect{\theta}) \quad \cdots \quad \vect{x}_P^{cT}(\tau, \vect{\theta})]^\mathrm{T}
\),
then Eq. \eqref{pcd_lin_dyn_1} can be expressed as a higher-dimensional deterministic linear dynamics system as
\begin{equation}
\label{pcd_lin_dyn}
\dot{\vect{x}}_G(\tau, \vect{\theta})
=
\vect{A}_G(\tau, \vect{\theta})\vect{x}_G(\tau, \vect{\theta}) + \vect{b}_G(\tau, \vect{\theta})
\end{equation}
where
\(\vect{b}_G(\tau, \vect{\theta}) = [\vect{b}_0^{cT}(\tau, \vect{\theta}) \quad \vect{b}_1^{cT}(\tau, \vect{\theta}) \quad \cdots \quad \vect{b}_P^{cT}(\tau, \vect{\theta})]^\mathrm{T}\);
\(\vect{A}_G(\tau, \vect{\theta}) = \sum_{i = 0}^{P}{\vect{\Phi}_i\otimes\vect{A}_i^c(\tau, \vect{\theta})}\);
\(\otimes\) denotes Kronecker product;
the matrix \(\vect{\Phi}_i\) is defined as
\begin{equation}
\vect{\Phi}_i =
\begin{bmatrix}
\dfrac{\langle\phi_i(\vect{\xi}_w)\phi_0(\vect{\xi}_w),\, \phi_0(\vect{\xi}_w)\rangle}{\langle\phi_0(\vect{\xi}_w),\, \phi_0(\vect{\xi}_w)\rangle}
%&
%\dfrac{\langle\phi_i(\vect{\xi})\phi_1(\vect{\xi}),\, \phi_0(\vect{\xi})\rangle}{\langle\phi_0(\vect{\xi}),\, \phi_0(\vect{\xi})\rangle}
& \cdots &
\dfrac{\langle\phi_i(\vect{\xi}_w)\phi_P(\vect{\xi}_w),\, \phi_0(\vect{\xi}_w)\rangle}{\langle\phi_0(\vect{\xi}_w),\, \phi_0(\vect{\xi}_w)\rangle}
%\\
%\dfrac{\langle\phi_i(\vect{\xi})\phi_0(\vect{\xi}),\, \phi_1(\vect{\xi})\rangle}{\langle\phi_1(\vect{\xi}),\, \phi_1(\vect{\xi})\rangle}
%&
%\dfrac{\langle\phi_i(\vect{\xi})\phi_1(\vect{\xi}),\, \phi_1(\vect{\xi})\rangle}{\langle\phi_1(\vect{\xi}),\, \phi_1(\vect{\xi})\rangle}
%& \cdots &
%\dfrac{\langle\phi_i(\vect{\xi})\phi_P(\vect{\xi}),\, \phi_1(\vect{\xi})\rangle}{\langle\phi_1(\vect{\xi}),\, \phi_1(\vect{\xi})\rangle}
\\
\vdots & \ddots & \vdots
\\
\dfrac{\langle\phi_i(\vect{\xi}_w)\phi_0(\vect{\xi}_w),\, \phi_P(\vect{\xi}_w)\rangle}{\langle\phi_P(\vect{\xi}_w),\, \phi_P(\vect{\xi}_w)\rangle}
%&
%\dfrac{\langle\phi_i(\vect{\xi})\phi_1(\vect{\xi}),\, \phi_P(\vect{\xi})\rangle}{\langle\phi_P(\vect{\xi}),\, \phi_P(\vect{\xi})\rangle}
& \cdots &
\dfrac{\langle\phi_i(\vect{\xi}_w)\phi_P(\vect{\xi}_w),\, \phi_P(\vect{\xi}_w)\rangle}{\langle\phi_P(\vect{\xi}_w),\, \phi_P(\vect{\xi}_w)\rangle}
\end{bmatrix}
\end{equation}

The initial value \(\vect{x}_G(\tau_c)\) of Eq. \eqref{pcd_lin_dyn} can be determined by
\begin{equation}
\label{xG0}
\vect{x}_G(\tau_\mathrm{c})
=
[\vect{x}_0^{cT}(\tau_\mathrm{c}) \quad \vect{x}_1^{cT}(\tau_\mathrm{c}) \quad \cdots \quad \vect{x}_P^{cT}(\tau_\mathrm{c})]^\mathrm{T}
,\quad
\sum_{k = 0}^{P}{\vect{x}_k^c(\tau_\mathrm{c})\phi_k(\vect{\xi}_w)}
=
\vect{x}_\mathrm{c}
\end{equation}

Then, the linear dynamics system as Eqs. \eqref{pcd_lin_dyn} and \eqref{xG0} is solved using Legendre-Guass-Radau (LGR) collocation method in a pseudospectral setting.
Define \(\varsigma = 2(\tau - \tau_\mathrm{c})/(1 - \tau_\mathrm{c}) - 1\) and approximate \(\vect{x}_G(\tau, \vect{\theta})\) using Lagrange interpolating polynomials as
\begin{equation}
\label{Lag-xG}
\vect{x}_G(\varsigma, \vect{\theta})
\approx
\sum_{i = 1}^{N_G}{\vect{x}_G(\varsigma_i, \vect{\theta})l_i(\varsigma_i)}
\end{equation}
where the Lagrange interpolation nodes contain the LGR integration points \(\varsigma_i(i = 1, 2, \cdots, N_G - 1)\) and the boundary node \(\varsigma_{N_G} = 1\).
Then Eqs. \eqref{pcd_lin_dyn} and \eqref{xG0} become
\begin{align}
\label{pcd_lin_dyn_dis}
&\sum_{i = 1}^{N_G}{D_{ji}\vect{x}_G(\varsigma_i, \vect{\theta})}
=
\frac{1 - \tau_c}{2}
[\vect{A}_G(\varsigma_j, \vect{\theta})\vect{x}_G(\varsigma_j, \vect{\theta}) + \vect{b}_G(\varsigma_j, \vect{\theta})], \quad j = 1, 2, \cdots, N_G - 1
\\
\label{xG0_dis}
&\vect{x}_G(\varsigma_1, \vect{\theta})
=
[\vect{x}_0^{cT}(\varsigma_1, \vect{\theta}) \quad \vect{x}_1^{cT}(\varsigma_1, \vect{\theta}) \quad \cdots \quad \vect{x}_P^{cT}(\varsigma_1, \vect{\theta})]^\mathrm{T}
\end{align}

Define
\(
\vect{X}_{Gs}(\vect{\theta})
=
[\vect{x}_G^{T}(\varsigma_1, \vect{\theta}) \quad \cdots \quad \vect{x}_G^{T}(\varsigma_{N_G}, \vect{\theta})]^\mathrm{T}
\), then
Eqs. \eqref{pcd_lin_dyn_dis} and \eqref{xG0_dis} can be expressed as
\begin{equation}
\label{pce_dyn_dis}
\vect{A}_{Gs}(\vect{\theta})\vect{X}_{Gs}(\vect{\theta}) = \vect{b}_{Gs}(\vect{\theta})
\end{equation}

Solving Eq. \eqref{pce_dyn_dis} yields \(\vect{X}_{Gs}(\vect{\theta}) = \vect{A}^{-1}_{Gs}(\vect{\theta})\vect{b}_{Gs}(\vect{\theta})\).
Taking each term of \(\vect{X}_{Gs}(\vect{\theta})\) and substituting it into Eq. \eqref{Lag-xG} yields \(\vect{x}_G(\varsigma, \vect{\theta})\), and the expansion coefficient \(\vect{x}_k^c(\tau, \vect{\theta})\) is obtained.

To this point, in Eq. \eqref{pce_x}, the closed-loop state is obtained as the analytical formulation of the random variable \(\vect{\xi}_w\) and the guidance parameter \(\vect{\theta}\).
As a result, the state trajectory dispersion can be predicted according to the probability density function of the random variable \(\vect{\xi}_w\).
To simplify the problem, this paper assumes that \(\vect{x}(\tau, \vect{\xi}_w, \vect{\theta})\) approximately follows normal distributions, and uses \(3\sigma\) principle to represent the trajectory dispersion with \(\SI{99.74}{\%}\) probability.
Using Gaussian integration, the mean vector and the variance vector of the state can be approximated as
\begin{align}
\label{E-x}
&\mathrm{E}[\vect{x}(\tau, \vect{\xi}_w, \vect{\theta})]
=
\sum_{i = 1}^{N_\xi}
{
w_i^\xi
\vect{x}(\tau, \vect{\xi}_{wi}, \vect{\theta})
}
\\
\label{V-x}
&\mathrm{V}[\vect{x}(\tau, \vect{\xi}_w, \vect{\theta})]
=
\sum_{i = 1}^{N_\xi}
{
w_i^\xi
\left\{
\vect{x}(\tau, \vect{\xi}_{wi}, \vect{\theta})
-
\mathrm{E}[\vect{x}(\tau, \vect{\xi}_w, \vect{\theta})]
\right\}
\odot
\left\{
\vect{x}(\tau, \vect{\xi}_{wi}, \vect{\theta})
-
\mathrm{E}[\vect{x}(\tau, \vect{\xi}_w, \vect{\theta})]
\right\}
}
\end{align}
where \(\mathrm{V}(\cdot)\) denote the variance vector;
\(w_i^\xi\) is the weight at the Gaussian integration point \(\vect{\xi}_{wi}\);
\(\odot\) is the Hadamard product, denoting element-wise multiplication.

As shown in Eqs. \eqref{E-x} and \eqref{V-x}, the state trajectory dispersion can be expressed using the combination of the mean value and the variance value, which is time-varying and varies with the guidance parameter.
In the next section, an real-time guidance parameter tuning law for the POFGL will be given to control the trajectory dispersion by satisfying the precision landing requirements and optimizing the performance index.

\section{Real-Time Guidance Parameter Tuning Law}
\label{Sec-5}

Based on the trajectory dispersion prediction, \cref{Sto-Param-Opt} can be transformed into a deterministic parameter optimization problem.
There are many approaches to achieving the parameter tuning for the POFGL, and from the standpoint of simplicity and computational efficiency, in this section, a real-time guidance parameter tuning law based on the gradient descent method is designed to achieve trajectory dispersion control.

In Eq. \eqref{pce_x}, the state \(\vect{x}(\tau, \vect{\xi}_w, \vect{\theta})\) has been represented as the analytical formulation of \(\vect{\xi}_w\) and \(\vect{\theta}\).
By substituting Eq. \eqref{pce_x} into Eqs. \eqref{simguid_law_u} and \eqref{simguid_law_a}, the guidance command \(\vect{u}(\tau, \vect{\xi}_w, \vect{\theta})\) and the total flight time \(\vect{a}(\tau, \vect{\xi}_w, \vect{\theta})\) can be represented as the analytical formulations of \(\vect{\xi}_w\) and \(\vect{\theta}\).
Consequently, based on the online trajectory dispersion prediction, \cref{Sto-Param-Opt} is approximated as a deterministic optimal parameter optimization problem as follows.

\begin{problem}
\label{Det-Param-Opt}
\normalfont
Trajectory Dispersion Control Problem Based on Online Dispersion Prediction
\begin{alignat}{2}
\label{Pst_J_s0_2}
&\mathop{\mathrm{minimize}}_{\vect{\theta} \in \vect{\Theta}}\quad
&&
J_{s}(\vect{\theta})
\\
&\,\text{subject to}\quad
&&
\label{Pst_constrain_c_2}
\vect{g}_c(\vect{\theta}) \leq \vect{0}
\\
\label{Pst_constrain_u_2}
&
&&
\vect{g}_u(\vect{\theta}) \leq \vect{0}
\end{alignat}
where ``\(\leq\)'' denotes that each element of the vector satisfies the less-than-equal relation;
\(
\vect{g}_u
=
[
\vect{g}_{ui}^\mathrm{T}
\quad \cdots \quad
\vect{g}_{uN_\varsigma}^\mathrm{T}
]^\mathrm{T}
\);
\begin{align}
\vect{g}_c
=
\left[
\begin{gathered}
\mathrm{E}[\vect{C}\vect{x}(1, \vect{\xi}_w, \vect{\theta})]
+ 3\sqrt{\mathrm{V}[\vect{C}\vect{x}(1, \vect{\xi}_w, \vect{\theta})]}
- \vect{c}_\mathrm{max}
\\
- \mathrm{E}[\vect{C}\vect{x}(1, \vect{\xi}_w, \vect{\theta})]
+ 3\sqrt{\mathrm{V}[\vect{C}\vect{x}(1, \vect{\xi}_w, \vect{\theta})]}
+ \vect{c}_\mathrm{min}
\end{gathered}
\right]
,\,\,
\vect{g}_{ui}
=
\left[
\begin{gathered}
\mathrm{E}[\vect{u}(\varsigma_i, \vect{\xi}_w, \vect{\theta})]
+ \sqrt{\mathrm{V}[\vect{u}(\varsigma_i, \vect{\xi}_w, \vect{\theta})]}
- \vect{u}_\mathrm{max}
\\
- \mathrm{E}[\vect{u}(\varsigma_i, \vect{\xi}_w, \vect{\theta})]
+ \sqrt{\mathrm{V}[\vect{u}(\varsigma_i, \vect{\xi}_w, \vect{\theta})]}
+ \vect{u}_\mathrm{min}
\end{gathered}
\right]
\end{align}
\end{problem}

In \cref{Det-Param-Opt}, using Gaussian integration, Eq. \eqref{Pst_J_s0_2} is approximated as the deterministic performance index as
\begin{equation}
\label{Js0-GI}
J_{s}(\vect{\theta}) =
\sum_{i = 1}^{N_\xi}
{
w_i^\xi
\left\{
-R_mm(1, \vect{\xi}_{wi}, \vect{\theta})
+
\sum_{j = 1}^{N_\varsigma}
{
\frac{1}{2}w_j^\varsigma
L_0[\vect{x}(\varsigma_j, \vect{\xi}_{wi}, \vect{\theta}), \vect{u}(\varsigma_j, \vect{\xi}_{wi}, \vect{\theta}), a(\varsigma_j, \vect{\xi}_{wi}, \vect{\theta})]
}
\right\}
}
\end{equation}
where
\(w_j^\varsigma\) is the weight at the Gaussian integration point \(\varsigma_j\).

Using \(3\sigma\) principle, Eq. \eqref{Pst_constrain_c_2} is derived by approximating Eq. \eqref{problem-sto-terminal-constraint} with \(\SI{99.74}{\%}\) probability as
\begin{equation}
\label{det-terminal-constraint}
\vect{c}_\mathrm{min}
\leq
\mathrm{E}[\vect{C}\vect{x}(1, \vect{\xi}_w, \vect{\theta})]
\pm3\sqrt{\mathrm{V}[\vect{C}\vect{x}(1, \vect{\xi}_w, \vect{\theta})]}
\leq
\vect{c}_\mathrm{max}
\end{equation}

Similarly, Eq. \eqref{Pst_constrain_u_2} is derived by approximating Eq. \eqref{problem-sto-control-constraint} with \(\SI{99.74}{\%}\) probability as
\begin{equation}
\label{det-control-constraint}
\vect{u}_\mathrm{min}
\leq
\mathrm{E}[\vect{u}(\varsigma_j, \vect{\xi}_w, \vect{\theta})]
\pm3\sqrt{\mathrm{V}[\vect{u}(\varsigma_j, \vect{\xi}_w, \vect{\theta})]}
\leq
\vect{u}_\mathrm{max}
,\quad
j = 1,\cdots,N_\varsigma
\end{equation}
where the mean and variance of the guidance command can be approximated via Gaussian integration respectively as
\begin{align}
&\mathrm{E}[\vect{u}(\varsigma_j, \vect{\xi}_w, \vect{\theta})]
=
\sum_{i = 1}^{N_\xi}
{
w_i^\xi
\vect{u}(\varsigma_j, \vect{\xi}_{wi}, \vect{\theta})
}
\\
&\mathrm{V}[\vect{u}(\varsigma_j, \vect{\xi}_w, \vect{\theta})]
=
\sum_{i = 1}^{N_\xi}
{
w_i^\xi
\left\{
\vect{u}(\varsigma_j, \vect{\xi}_{wi}, \vect{\theta})
-
\mathrm{E}[\vect{u}(\varsigma_j, \vect{\xi}_w, \vect{\theta})]
\right\}
\odot
\left\{
\vect{u}(\varsigma_j, \vect{\xi}_{wi}, \vect{\theta})
-
\mathrm{E}[\vect{u}(\varsigma_j, \vect{\xi}_w, \vect{\theta})]
\right\}
}
\end{align}

To efficiently solve \cref{Det-Param-Opt}, the penalty function method is used to transform \cref{Det-Param-Opt} into an unconstrained parameter optimization problem, denoted as
\begin{equation}
\label{Js}
\mathrm{minimize}\quad
J_{t}(\vect{\theta}) = J_{s}(\vect{\theta})
+
\sum_{i}{F_\mathrm{Pen}[g_{ci}(\vect{\theta})]}
+
\sum_{i}{F_\mathrm{Pen}[g_{ui}(\vect{\theta})]}
\end{equation}
where \(g_{ui}(\vect{\theta})\) denotes each row in the vector \(\vect{g}_u(\vect{\theta})\);
\(g_{ci}(\vect{\theta})\) denotes each row in the vector \(\vect{g}_c(\vect{\theta})\);
\(F_\mathrm{Pen}(\cdot)\) is the penalty function, and this paper adopts the well-known smoothed Hinge penalty function, denoted as
\begin{equation}
F_\mathrm{Pen}(g)
=
\frac{1}{\sigma_\mathrm{Pen}}
\ln{[1 + \exp{(\sigma_\mathrm{Pen}g)}]}
\end{equation}
where \(\sigma_\mathrm{Pen}\) is smoothing parameter.

Then, to achieve trajectory dispersion control, a quadratic weighting parameter tuning law based on gradient descent method is designed as
\begin{equation}
\label{learning-law-online}
\vect{\theta}(\tau_{c + 1}) = \vect{\Pi}_\Theta
\left[
\vect{\theta}(\tau_c)
-
\gamma
\left.
\left(\frac{\partial J_t}{\partial\vect{\theta}}\right)^\mathrm{T}
\right|_{\vect{\theta} = \vect{\theta}(\tau_c)}
\right]
,\quad
\vect{\theta}(0) = \vect{\theta}_\mathrm{init}
\end{equation}
where
\(\vect{\Pi}_\Theta\) denotes the Euclidean projection of \(\vect{\theta}\) onto \(\vect{\Theta}\);
\(\vect{\theta}(\tau_c)\) is the parameter at the time \(\tau_c\);
\(\vect{\theta}(\tau_{c + 1})\) is the parameter at the time \(\tau_{c + 1} = \tau_c + \Delta\tau_\mathrm{learn}\);
\(\Delta\tau_\mathrm{learn}\) is the online parameter tuning period;
\(\gamma^{(i)} \geq 0\) is the descent step size;
\(\vect{\theta}_\mathrm{init}\) is the initial value of guidance parameter.

The initial value \(\vect{\theta}_\mathrm{init}\) can be obtained offline using a random search method based on Latin Hypercube Sampling (LHS), denoted as
\begin{equation}
\label{Random-Search}
\vect{\theta}_\mathrm{init}
=
\mathrm{arg}
\mathop{\mathrm{min}}\limits_{\vect{\theta} \in \vect{\Theta}_{\mathrm{LHS}}}
J_{t}(\vect{\theta})
\end{equation}
where \(\vect{\Theta}_{\mathrm{LHS}}\) is the set of parameter samples formed by the LHS.
It is worth mentioning that, in the offline guidance parameter optimization, the initial state is unknown and can be described as
\(
\vect{x}_0 = \vect{\mu}_0 + \vect{\xi}_0
\),
where
\(\vect{\mu}_0\) is the mean vector of the initial state;
\(\vect{\xi}_0\) is the random vector related to the initial state.
As a result, Eqs. \eqref{pce_x} and \eqref{xG0} becomes
\begin{equation}
\label{pce_x_offline}
\vect{x}(\tau, \vect{\xi}, \vect{\theta})
=
\sum_{k = 0}^{P_\mathrm{off}}{\vect{x}_k^c(\tau, \vect{\theta})\phi_k(\vect{\xi})}
,\quad
\sum_{k = 0}^{P_\mathrm{off}}{\vect{x}_k^c(0)\phi_k(\vect{\xi})}
= \vect{\mu}_0 + \vect{\xi}_0
\end{equation}
where \(\vect{\xi} = [\vect{\xi}_w^\mathrm{T} \quad \vect{\xi}_0^\mathrm{T}]\);
\(P_\mathrm{off}\) is the number of truncated terms in the offline parameter tuning.

The proposed trajectory dispersion prediction method establishes an explicit relationship between random variables and guidance performance using gPC expansion, and thanks to the gradient descent method, the proposed parameter tuning law is simple and adaptable for online implementation, especially for the missions with nonlinear dynamics.

\section{Numerical Verification}
\label{Sec-6}

For numerical demonstration, the rocket parameters and nominal initial state are shown in Ref. \cite{Paper-Optimal}.
The processor of the test computer is an Advanced Micro Devices (AMD) Ryzen 7 6800H with a 3.2 GHz clock speed.
The matrices \(\vect{R}_0\) and \(\vect{R}_1\) are set as \(\vect{R}_0 = \mathrm{diag}(100,\,100,\,100)\) and \(\vect{R}_1 = \mathrm{diag}(1,\,1,\,1,\,1,\,1,\,1,\,1000,\,1000)\).
The minimum control allowed is \(\vect{u}_\mathrm{min} = [0.6,\,-\SI{5}{\degree/s},\,-\SI{10}{\degree/s}]^\mathrm{T}\) and the maximum control allowed is \(\vect{u}_\mathrm{max} = [1.0,\,+\SI{5}{\degree/s},\,+\SI{10}{\degree/s}]^\mathrm{T}\).
The random variables \(\xi_{T}\), \(\xi_{\varphi}\), \(\xi_{\psi}\), \(\xi_{Cx}\), \(\xi_{Cy}\), \(\xi_{Cz}\) and \(\xi_{\rho}\) follow zero-mean normal distributions, and their \(3\sigma\) values are respectively: \(\xi_{T}^{(3\sigma)} = \SI{3}{\%}\), \(\xi_{\varphi}^{(3\sigma)} = \SI{0.5}{\degree}\), \(\xi_{\psi}^{(3\sigma)} = \SI{0.5}{\degree}\), \(\xi_{Cx}^{(3\sigma)} = \SI{50}{\%}\), \(\xi_{Cy}^{(3\sigma)} = \SI{50}{\%}\), \(\xi_{Cz}^{(3\sigma)} = \SI{50}{\%}\), and \(\xi_{\rho}^{(3\sigma)} = \SI{30}{\%}\).
The wind velocity is polynomially fitted at altitudes \(h_{1} = \SI{0}{km}\), \(h_{2} = \SI{1.5}{km}\), and \(h_{3} = \SI{3.5}{km}\);
the wind velocities follow zero-mean normal distributions, and their \(3\sigma\) values are respectively: \(\xi_{V1}^{(3\sigma)} = \SI{30}{m/s}\), \(\xi_{V2}^{(3\sigma)} = \SI{60}{m/s}\), and \(\xi_{V3}^{(3\sigma)} = \SI{40}{m/s}\).
The wind direction is uniformly distributed between \([-\SI{90}{\degree}, +\SI{90}{\degree}]\).
The mean vector of the initial state is equal to the nominal initial state;
each of the elements in the random vector \(\vect{\xi}_0\) follows a zero-mean normal distribution, and their \(3\sigma\) values are respectively: \(\xi_{rx0}^{(3\sigma)} = \SI{300}{m}\), \(\xi_{ry0}^{(3\sigma)} = \SI{300}{m}\), \(\xi_{rz0}^{(3\sigma)} = \SI{300}{m}\), \(\xi_{vx0}^{(3\sigma)} = \SI{15}{m/s}\), \(\xi_{vy0}^{(3\sigma)} = \SI{15}{m/s}\), \(\xi_{vz0}^{(3\sigma)} = \SI{15}{m/s}\), \(\xi_{\varphi0}^{(3\sigma)} = \SI{5}{\degree}\), and \(\xi_{\psi0}^{(3\sigma)} = \SI{5}{\degree}\).
The polynomial time-varying weight matrices have degrees of freedom \(M = 3\).
The guidance period is \(\SI{10}{ms}\) and the parameter tuning period is \(\SI{100}{ms}\).
In the following simulations, the effectiveness of the trajectory dispersion control method is validated, and then the landing accuracy is analyzed.

\subsection{Effectiveness Analysis of Trajectory Dispersion Control}

To validate the effectiveness of the proposed trajectory dispersion control method, three simulation cases are carried out:
in Case 0, the guidance parameter is tuned offline considering the initial state dispersion, and the desired landing accuracy for terminal position, velocity, and attitude angle are respectively \(\SI{5}{m}\), \(\SI{2}{m/s}\) and \(\SI{2}{\degree}\);
in Case 1, the offline tuned guidance parameter is used to predict the trajectory dispersion at the initial time; 
in Case 2, the guidance parameter is tuned at the initial time by using the gradient descent algorithm for 100 times, and the desired landing accuracy for terminal position, velocity, and attitude angle are respectively \(\SI{1}{m}\), \(\SI{0.5}{m/s}\) and \(\SI{1}{\degree}\).

\begin{figure*}[!h]
        \centering
    \begin{subfigure}[h]{0.44\linewidth}
        \centering
        \includegraphics[width=\textwidth]{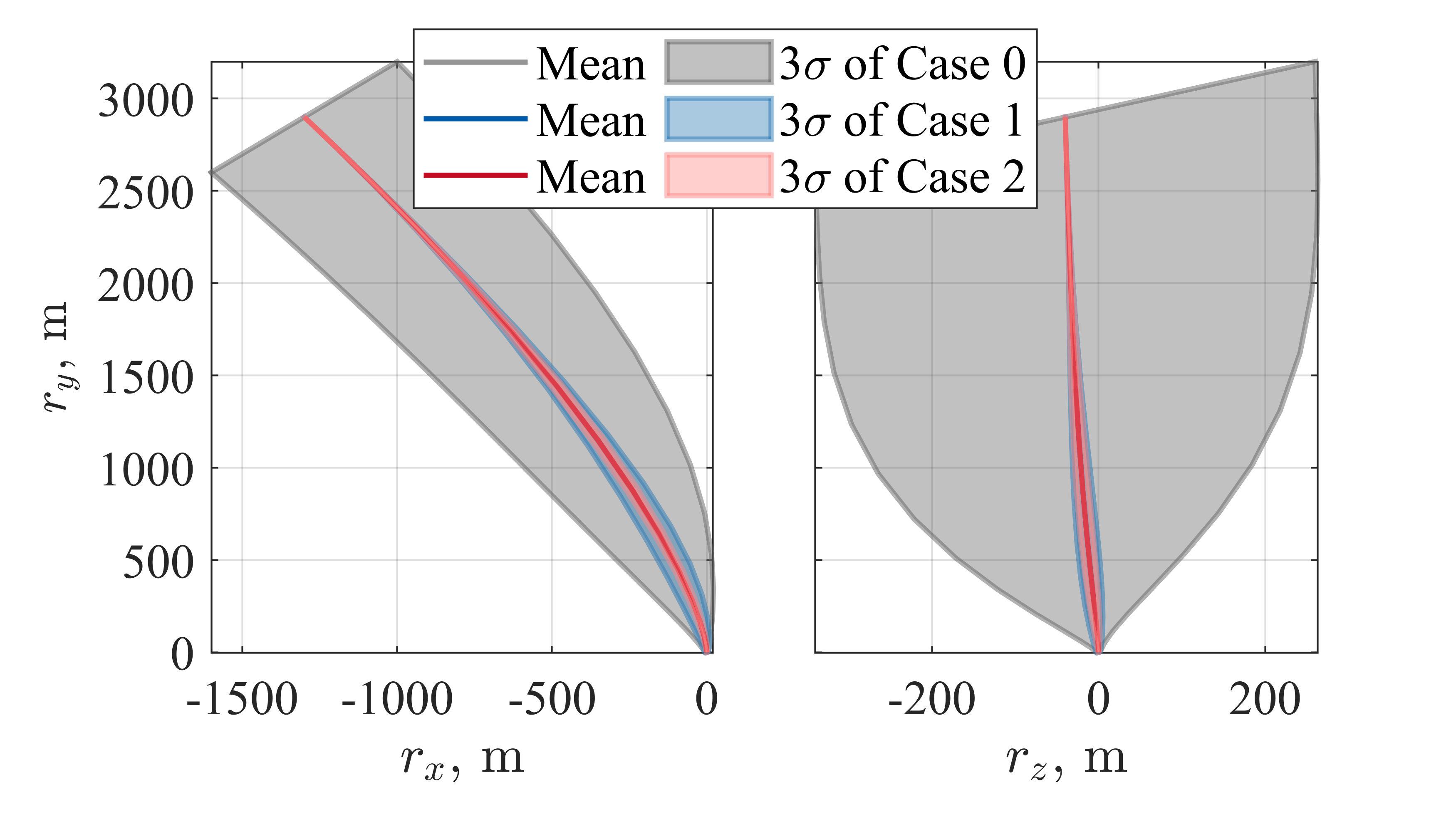}
        \caption{Position trajectory dispersions.}
    \end{subfigure}
	 \begin{subfigure}[h]{0.44\linewidth}
        \centering
        \includegraphics[width=\textwidth]{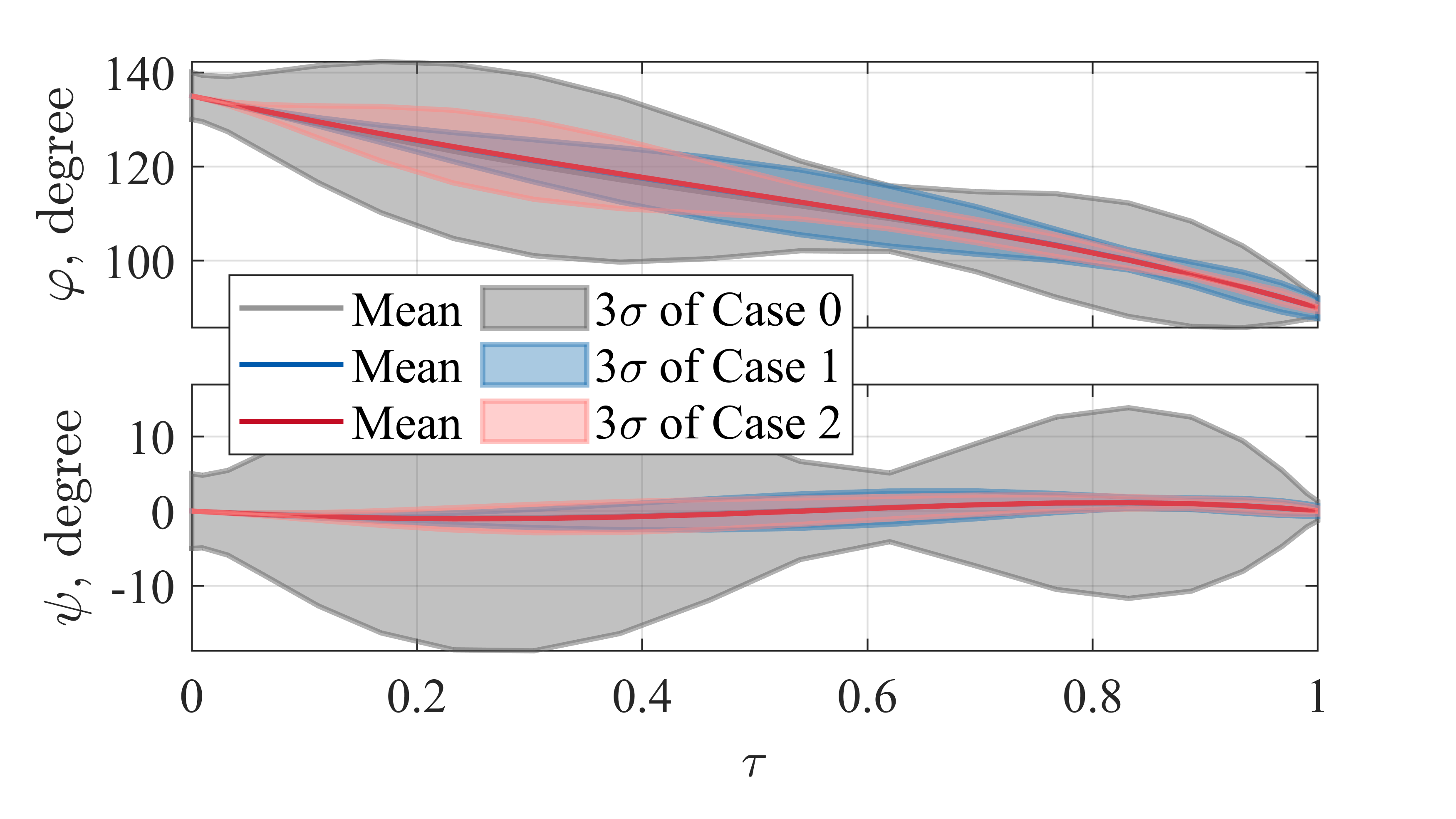}
        \caption{Attitude angle trajectory dispersions.}
    \end{subfigure}
\\
	\begin{subfigure}[h]{0.44\linewidth}
        \centering
        \includegraphics[width=\textwidth]{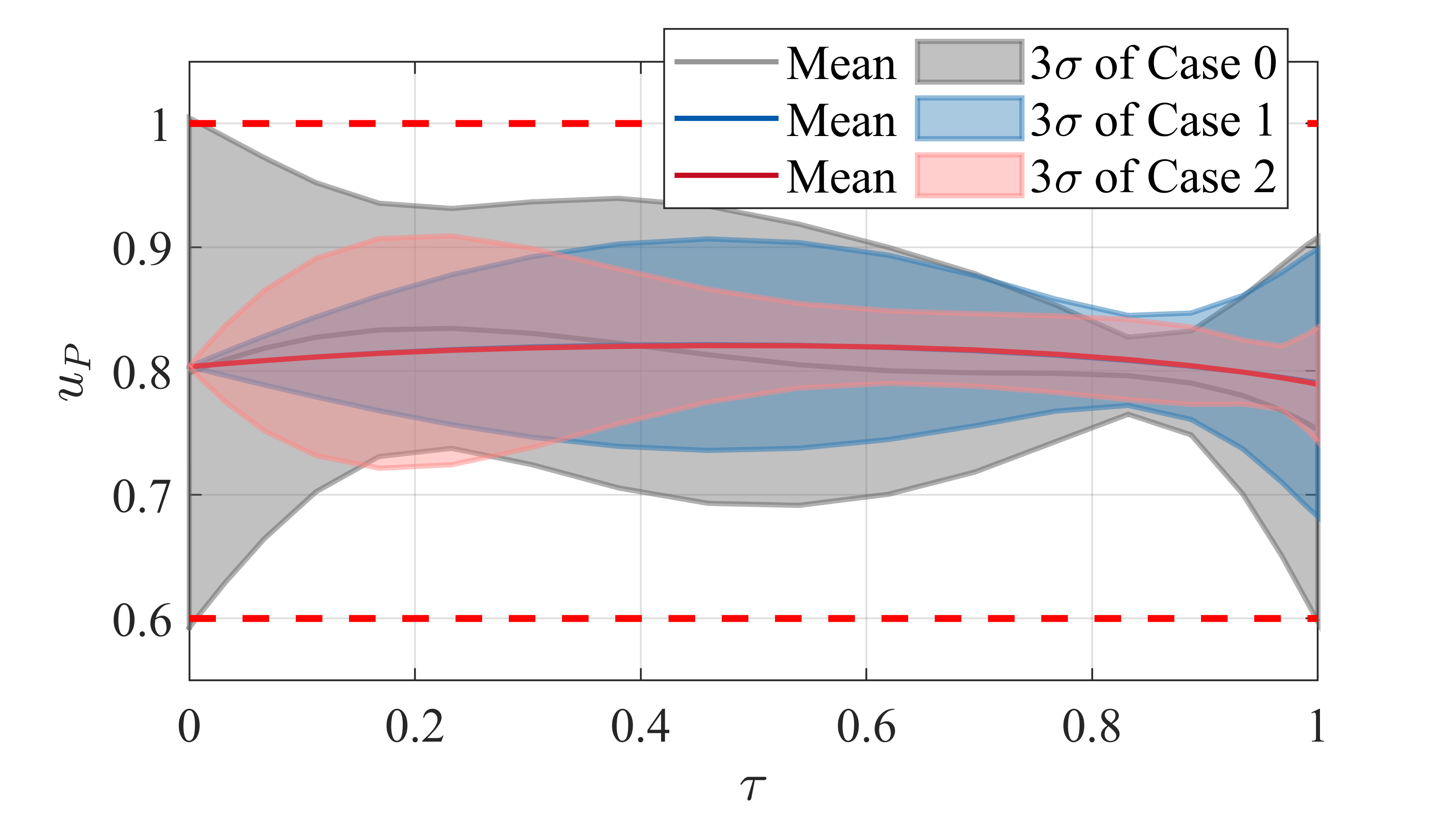}
        \caption{Engine throttling ratio trajectory dispersions.}
        \label{fig-sim-1-u}
    \end{subfigure}
    \begin{subfigure}[h]{0.44\linewidth}
        \centering
        \includegraphics[width=\textwidth]{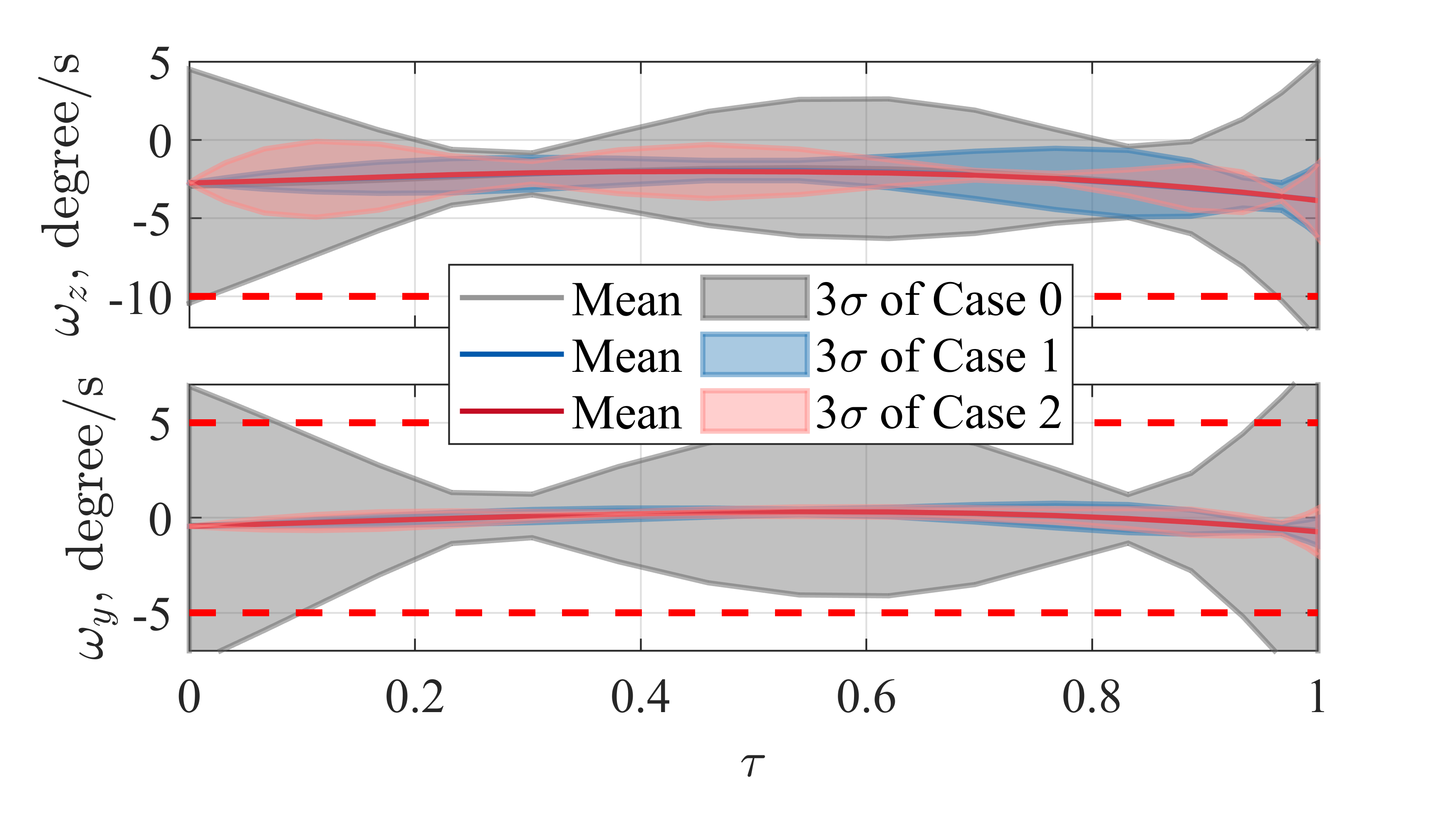}
        \caption{Attitude angular rate trajectory dispersions.}
        \label{fig-sim-1-w}
    \end{subfigure}
    \caption{Trajectory dispersion prediction results of ``Case 0'', ``Case 1'' and ``Case 2''.}
    \label{fig-sim-1}
\end{figure*}

\begin{figure*}[!h]
    \centering
    \includegraphics[width=0.88\linewidth]{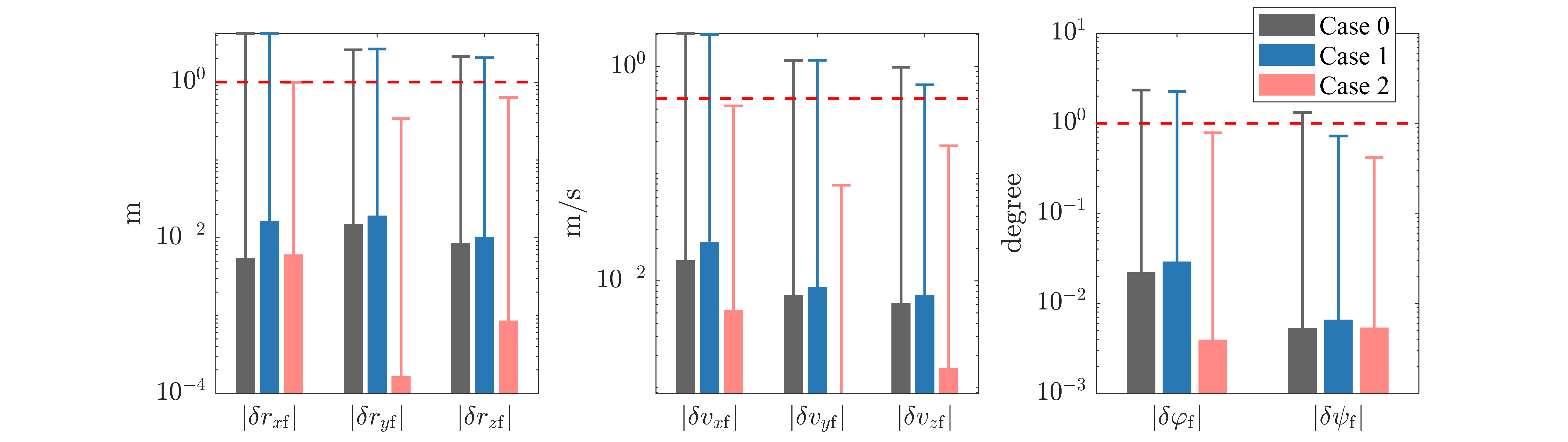}
    \caption{Terminal landing error bars.}
    \label{fig-sim-1-bar}
\end{figure*}

\begin{figure}[!h]
    \centering
    \begin{minipage}[t]{0.44\linewidth}
        \centering
        \includegraphics[width=1.0\textwidth]{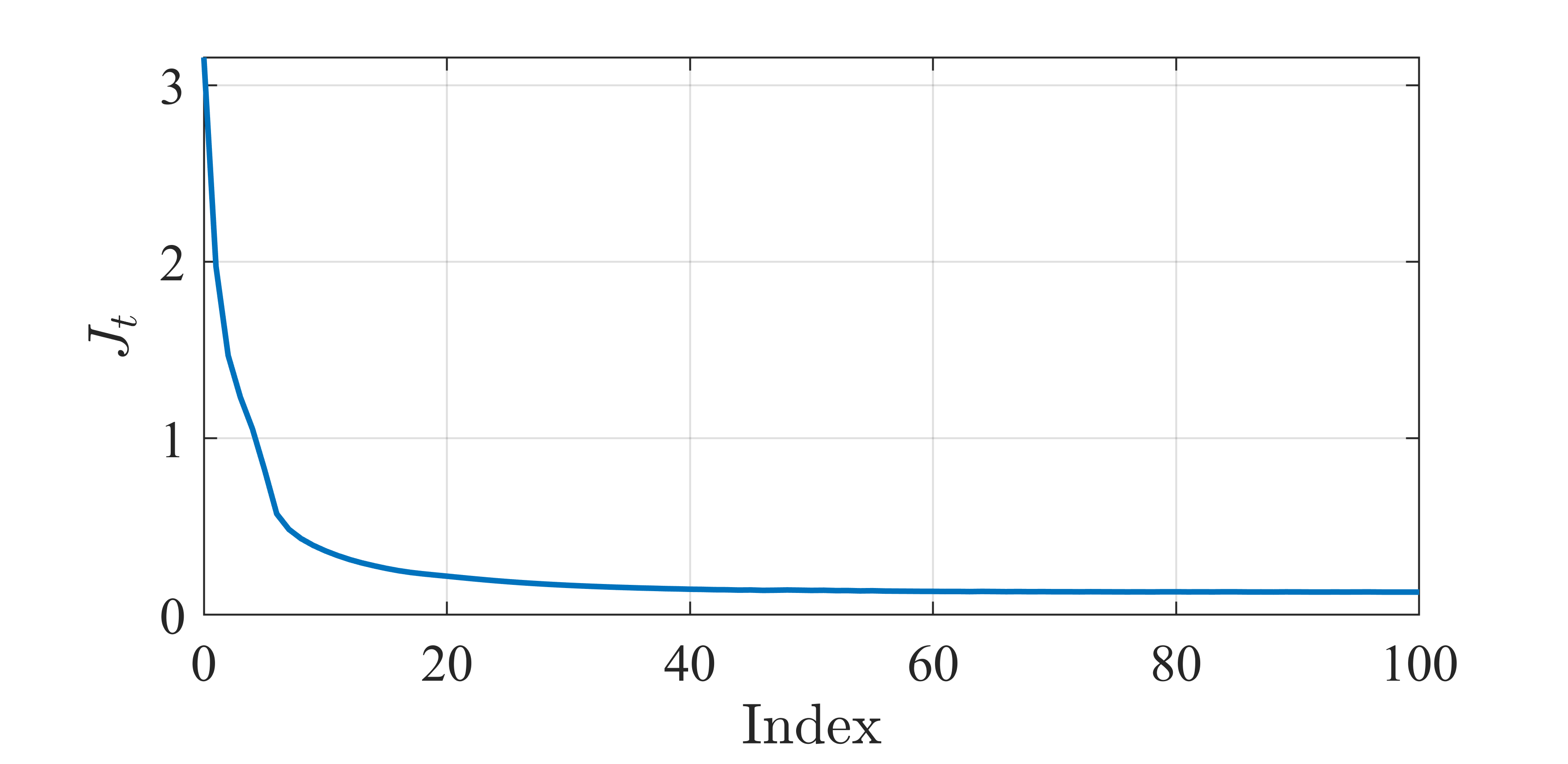}
    	\caption{Performance index profile.}
    	\label{fig-sim1-J}
    \end{minipage}%
    \begin{minipage}[t]{0.44\linewidth}
        \centering
        \includegraphics[width=1.0\textwidth]{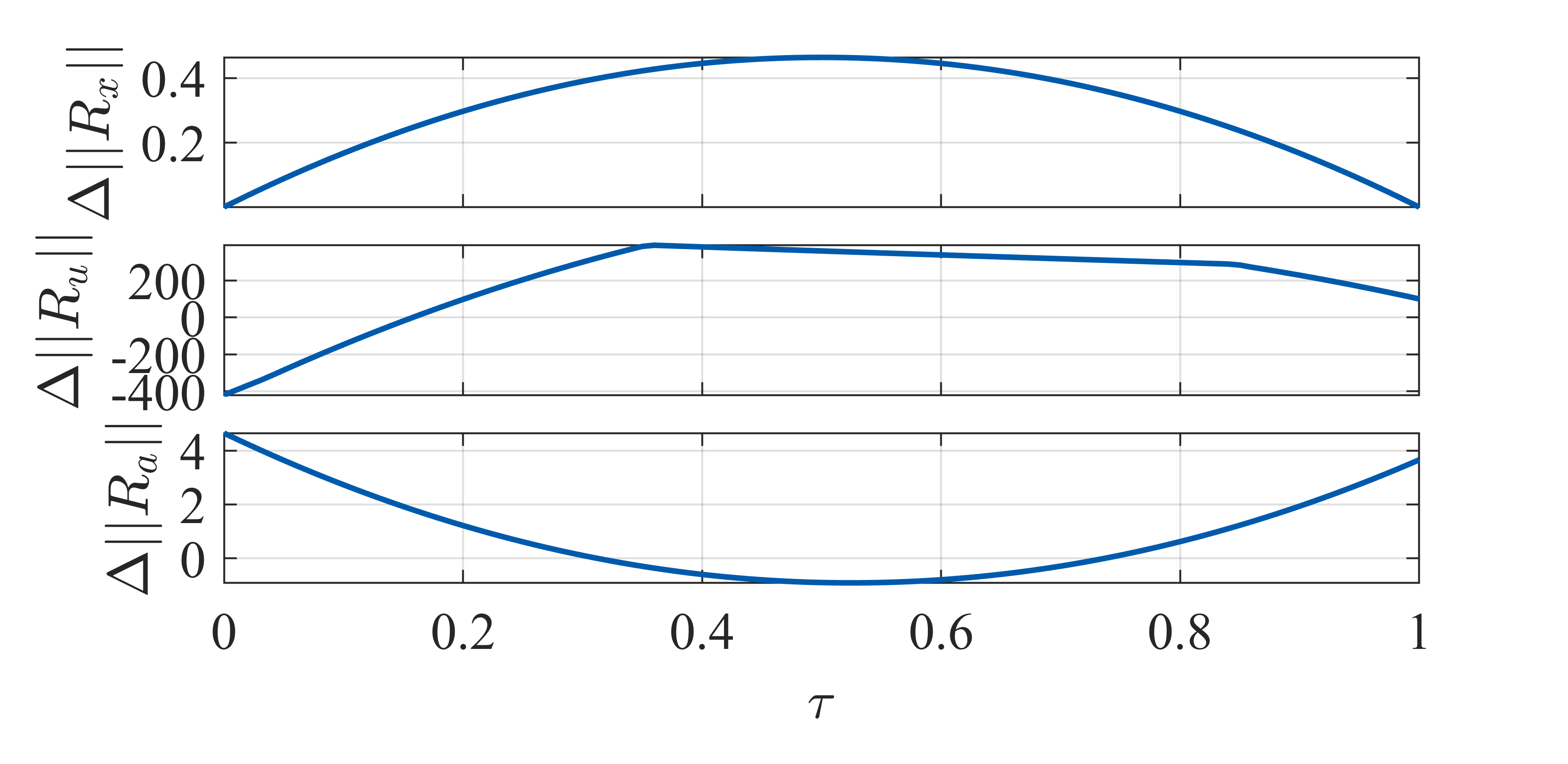}
        \caption{Guidance parameter profiles.}
    	\label{fig-sim1-theta}
    \end{minipage}
\end{figure}

\cref{fig-sim-1} gives the trajectory dispersion prediction results of ``Case 0'', ``Case 1'' and ``Case 2''.
It can be seen that, due to the uncertainty of the initial state in ``Case 0'', the trajectory dispersions are large, and the attitude angle rates shows a failure to satisfy the probabilistic constraints.
With the initial state determined in ``Case 1'', the trajectory dispersions are significantly reduced.
In ``Case 2'', compared with ``Case 1'', the proposed method intuitively optimizes the shape of the trajectory dispersion: the position trajectory dispersion decreases, and the trajectory dispersions of attitude angle, throttling rate and attitude angle rate increase in the initial portion and decrease in the later portion.
The throttling rate and attitude angle rate satisfy the probabilistic constraints.
\cref{fig-sim-1-bar} gives the terminal landing error bars of three cases,
where the red dashed lines denote the desired accuracy requirements.
\cref{fig-sim1-J} gives the profile of the performance index \(J_t\) in the gradient descent algorithm.
It can be seen that the guidance accuracy can meet the desired requirements, and the performance index decreases rapidly and converges around 40 times.
\cref{fig-sim1-theta} gives the difference between the norms of the time-varying weight matrices before and after the parameter tuning, and the difference of the norm of terminal weight matrix is \(\Delta\norm{\vect{R}_\mathrm{f}}_2 = 16.19\).
It can be seen that the state weight matrix \(\vect{R}_x\) and terminal weight matrix \(\vect{R}_\mathrm{f}\) increase to reduce the state trajectory dispersion and improve the landing accuracy.
The control weight matrix decreases in the initial portion and increases in the later portion, so that the guidance command dispersions increase in the initial portion and decrease in the later portion as shown in \cref{fig-sim-1-u} and \cref{fig-sim-1-w}.
The above results show that the proposed method improves the landing accuracy and performance index by directly shaping the trajectory dispersion in real time.


\begin{figure*}[!h]
        \centering
    \begin{subfigure}[h]{0.44\linewidth}
        \centering
        \includegraphics[width=\textwidth]{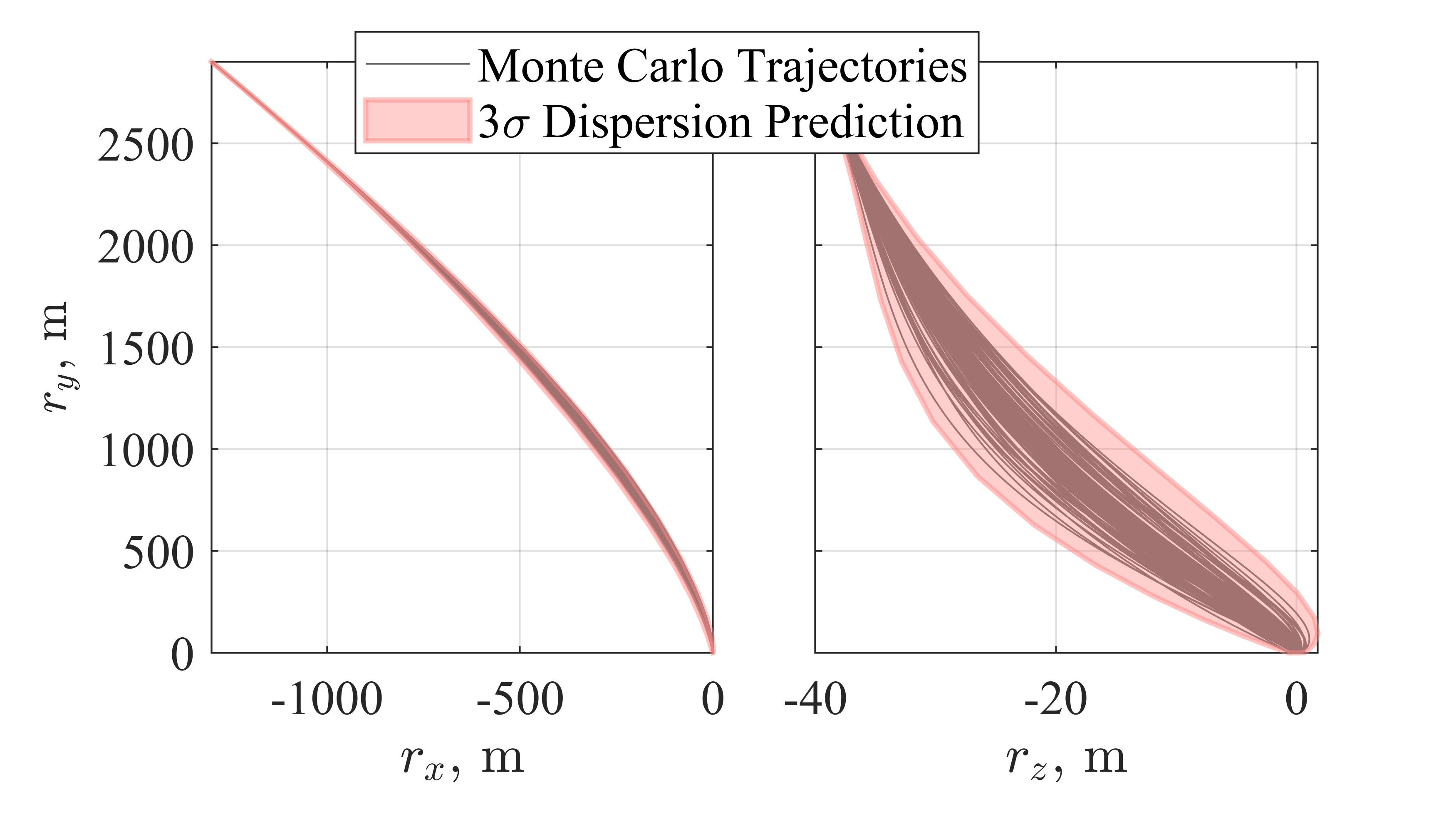}
        \caption{Position trajectory dispersions.}
    \end{subfigure}
	 \begin{subfigure}[h]{0.44\linewidth}
        \centering
        \includegraphics[width=\textwidth]{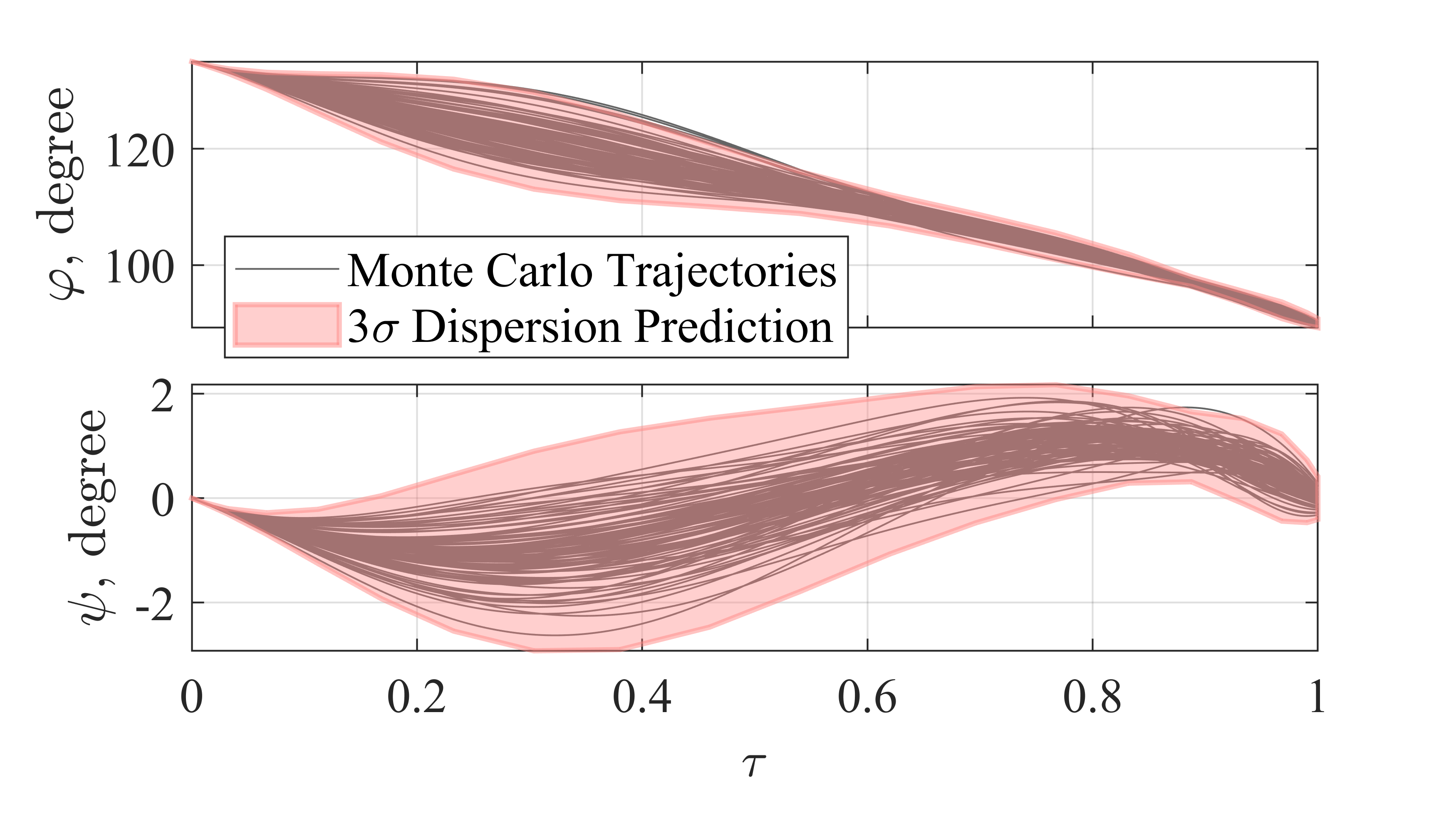}
        \caption{Attitude angle trajectory dispersions.}
    \end{subfigure}
\\
	\begin{subfigure}[h]{0.44\linewidth}
        \centering
        \includegraphics[width=\textwidth]{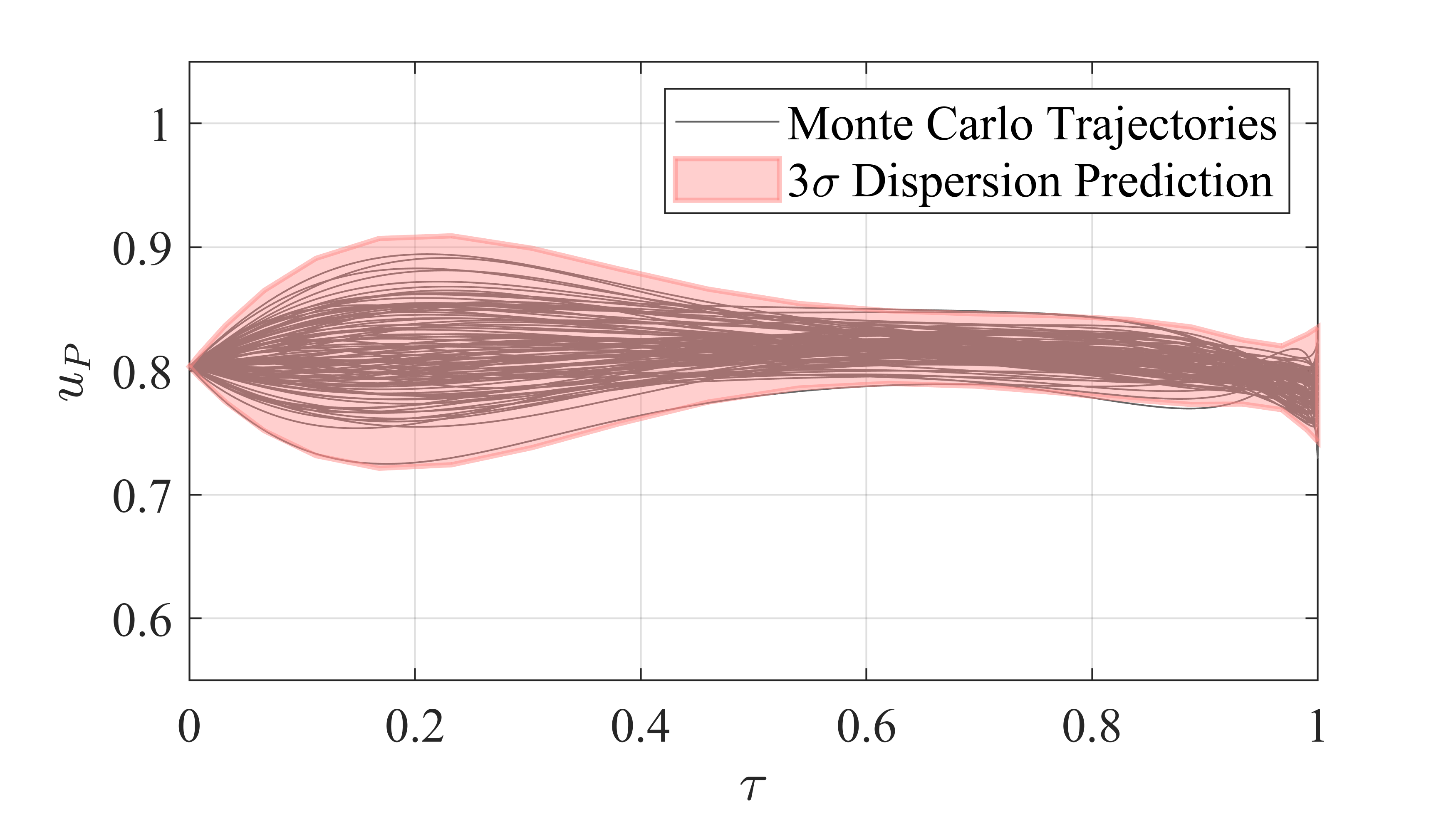}
        \caption{Engine throttling ratio trajectory dispersions.}
    \end{subfigure}
    \begin{subfigure}[h]{0.44\linewidth}
        \centering
        \includegraphics[width=\textwidth]{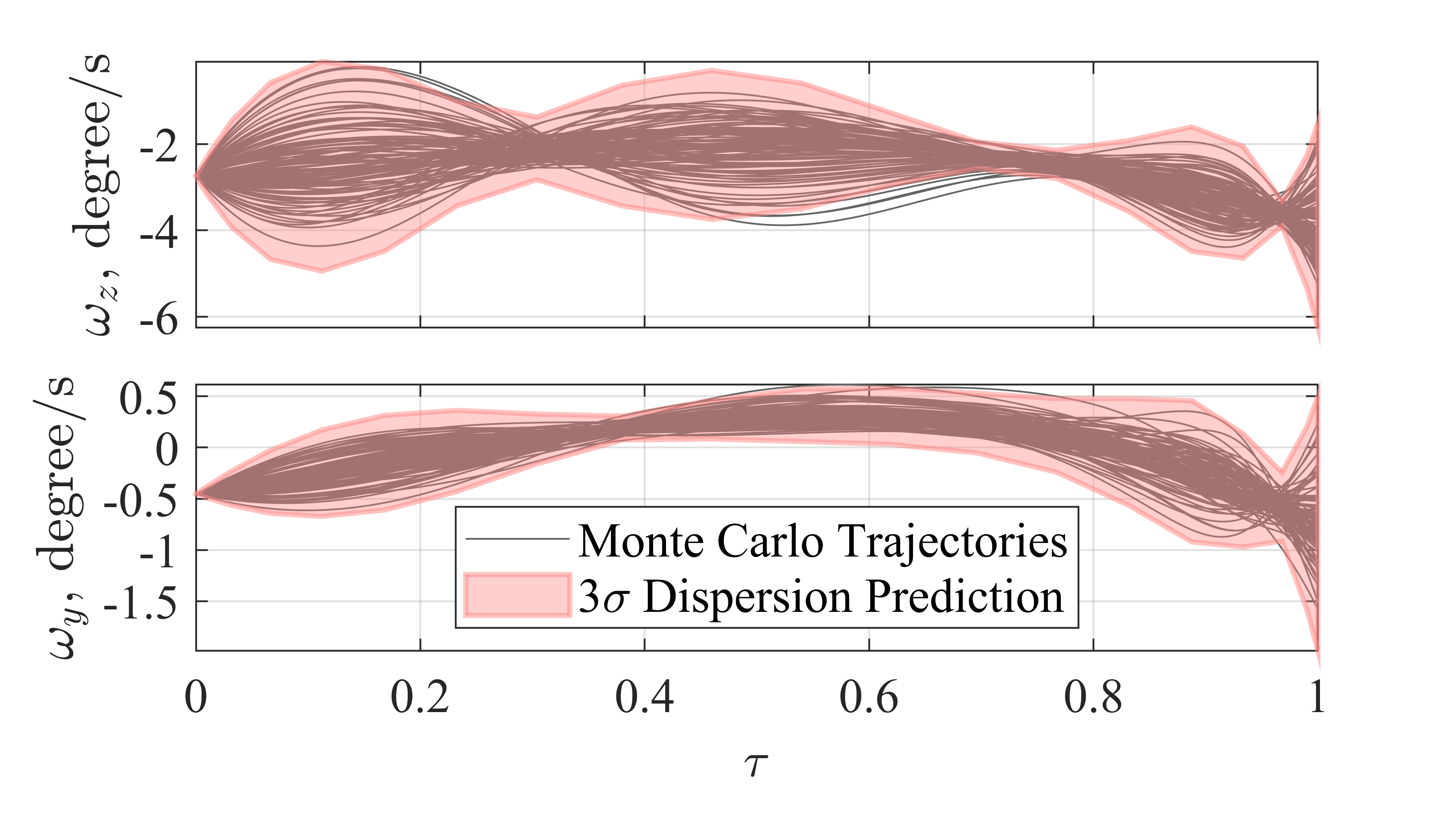}
        \caption{Attitude angular rate trajectory dispersions.}
    \end{subfigure}
    \caption{Comparison between online trajectory dispersion prediction and Monte Carlo simulation.}
    \label{fig-sim-1-Monte}
\end{figure*}

In order to validate the accuracy of the online trajectory dispersion prediction, \cref{fig-sim-1-Monte} gives the comparison between the predicted dispersion result and the Monte Carlo result consisting of 1000 simulations.
It is observed that the online trajectory dispersion prediction has high accuracy and the trajectory dispersion of \(3\sigma\) can cover almost the entire Monte Carlo simulation trajectories,
indicating that the proposed online trajectory dispersion prediction method is consistent with the realistic Monte Carlo design.
The average value of computational time consumed for a single trajectory dispersion prediction is \(\SI{10.06}{ms}\), the maximum value is \(\SI{12.32}{ms}\), and the minimum value is \(\SI{9.59}{ms}\), all of which are less than the parameter tuning period, satisfying the real-time computational requirement.

\subsection{Landing Accuracy Analysis}

\begin{figure*}[!h]
        \centering
    \begin{subfigure}[h]{0.44\linewidth}
        \centering
        \includegraphics[width=\textwidth]{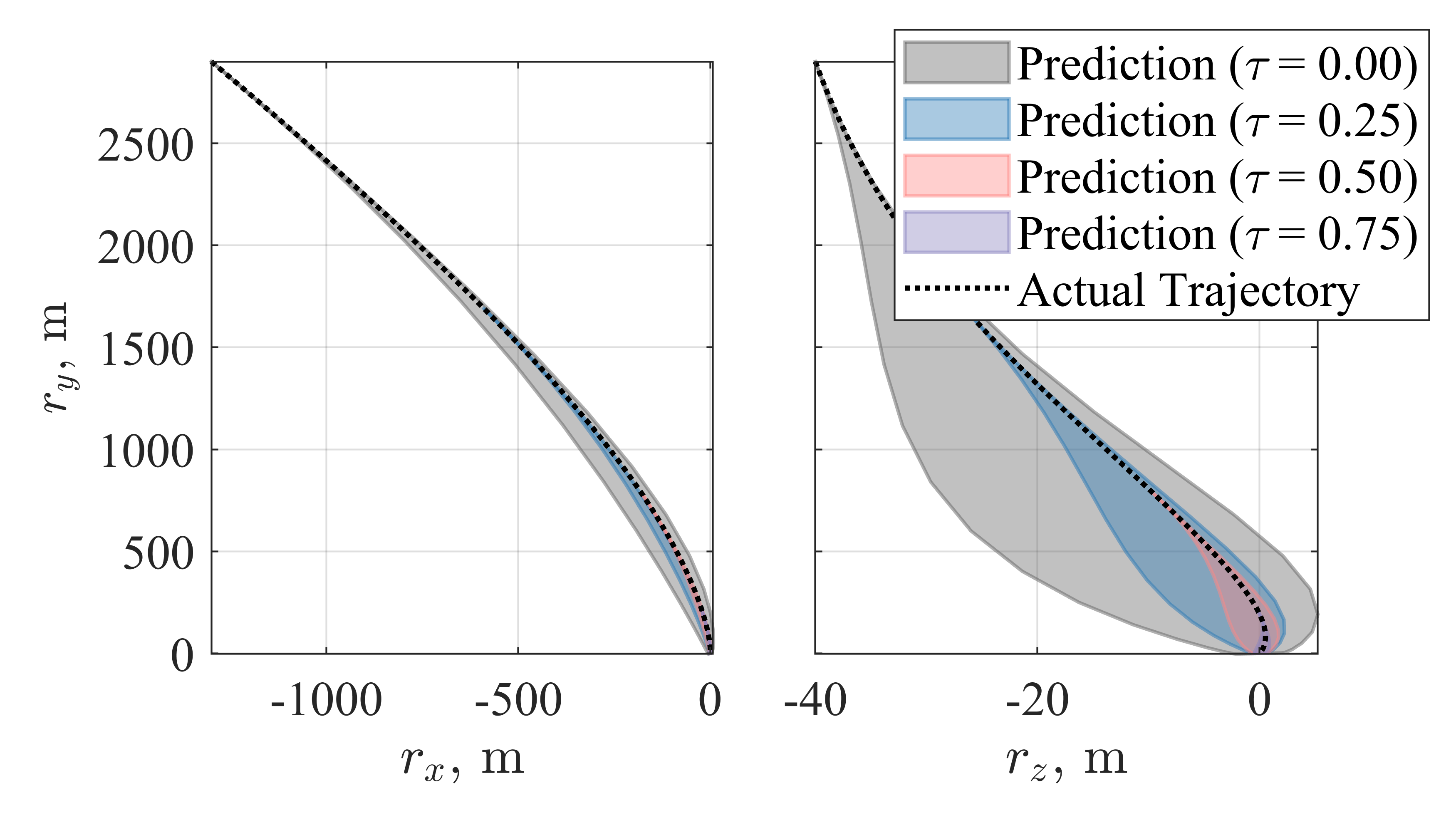}
        \caption{Position trajectory dispersions.}
    \end{subfigure}
	 \begin{subfigure}[h]{0.44\linewidth}
        \centering
        \includegraphics[width=\textwidth]{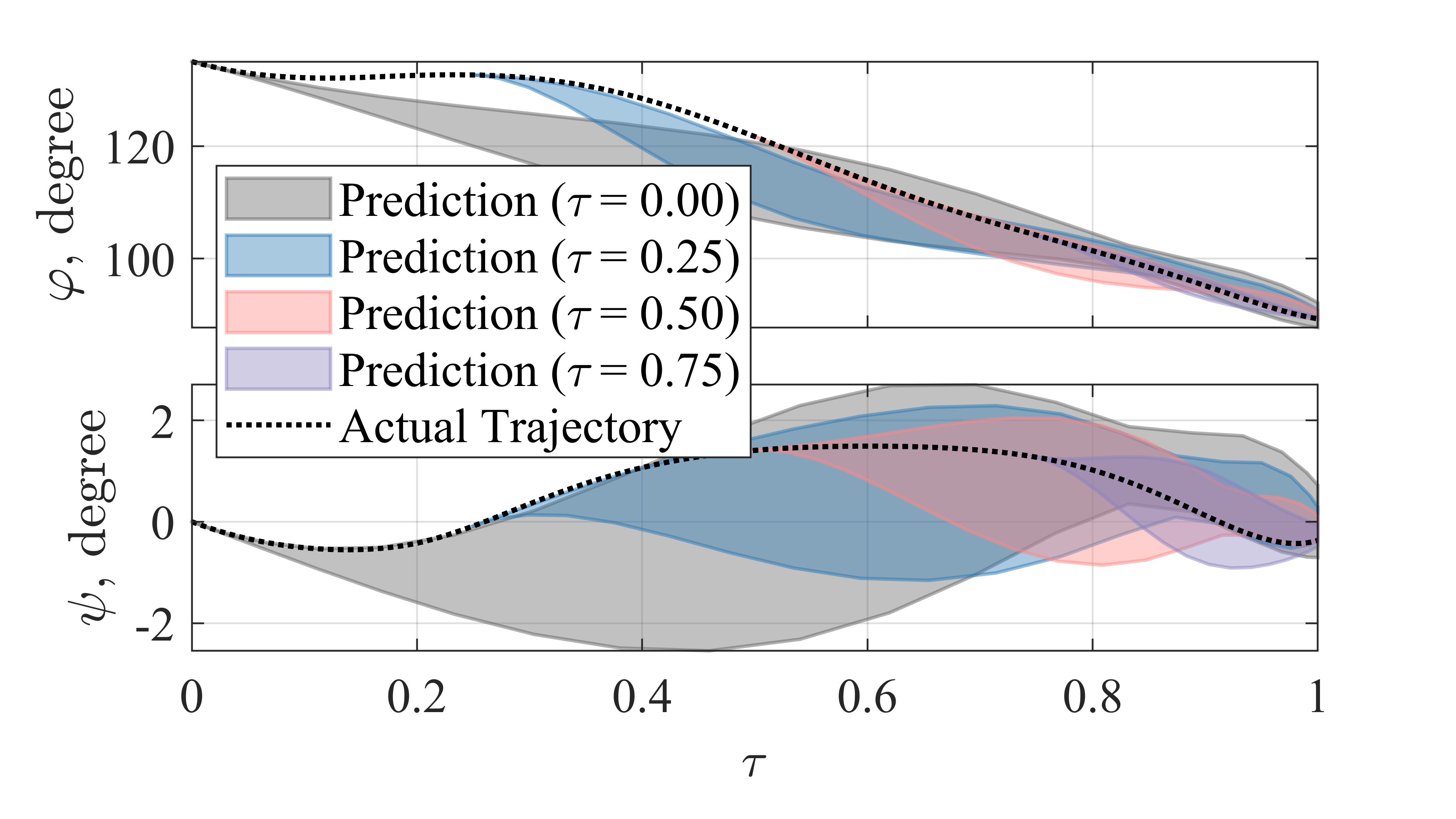}
        \caption{Attitude angle trajectory dispersions.}
    \end{subfigure}
\\
	\begin{subfigure}[h]{0.44\linewidth}
        \centering
        \includegraphics[width=\textwidth]{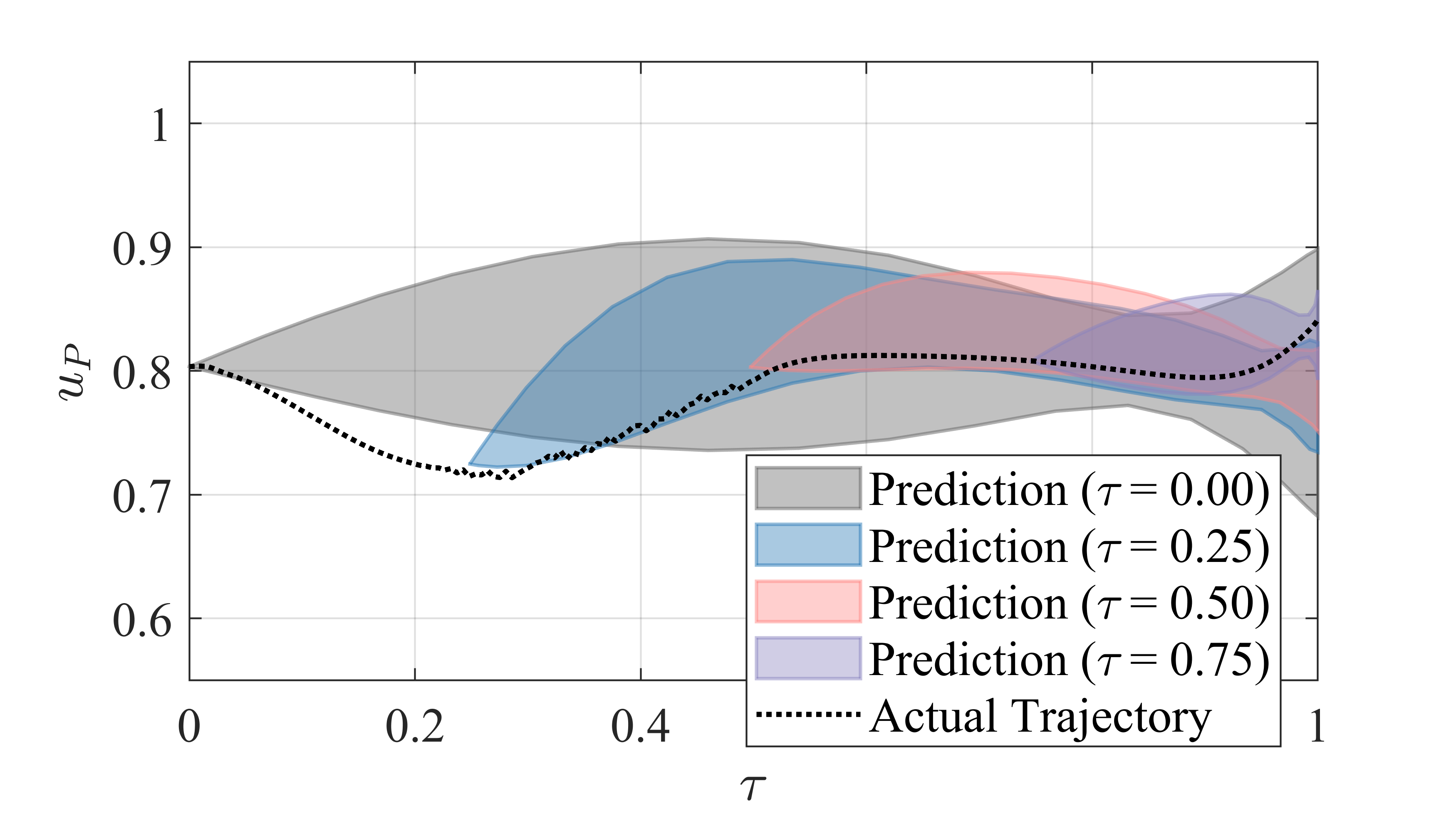}
        \caption{Engine throttling ratio trajectory dispersions.}
    \end{subfigure}
    \begin{subfigure}[h]{0.44\linewidth}
        \centering
        \includegraphics[width=\textwidth]{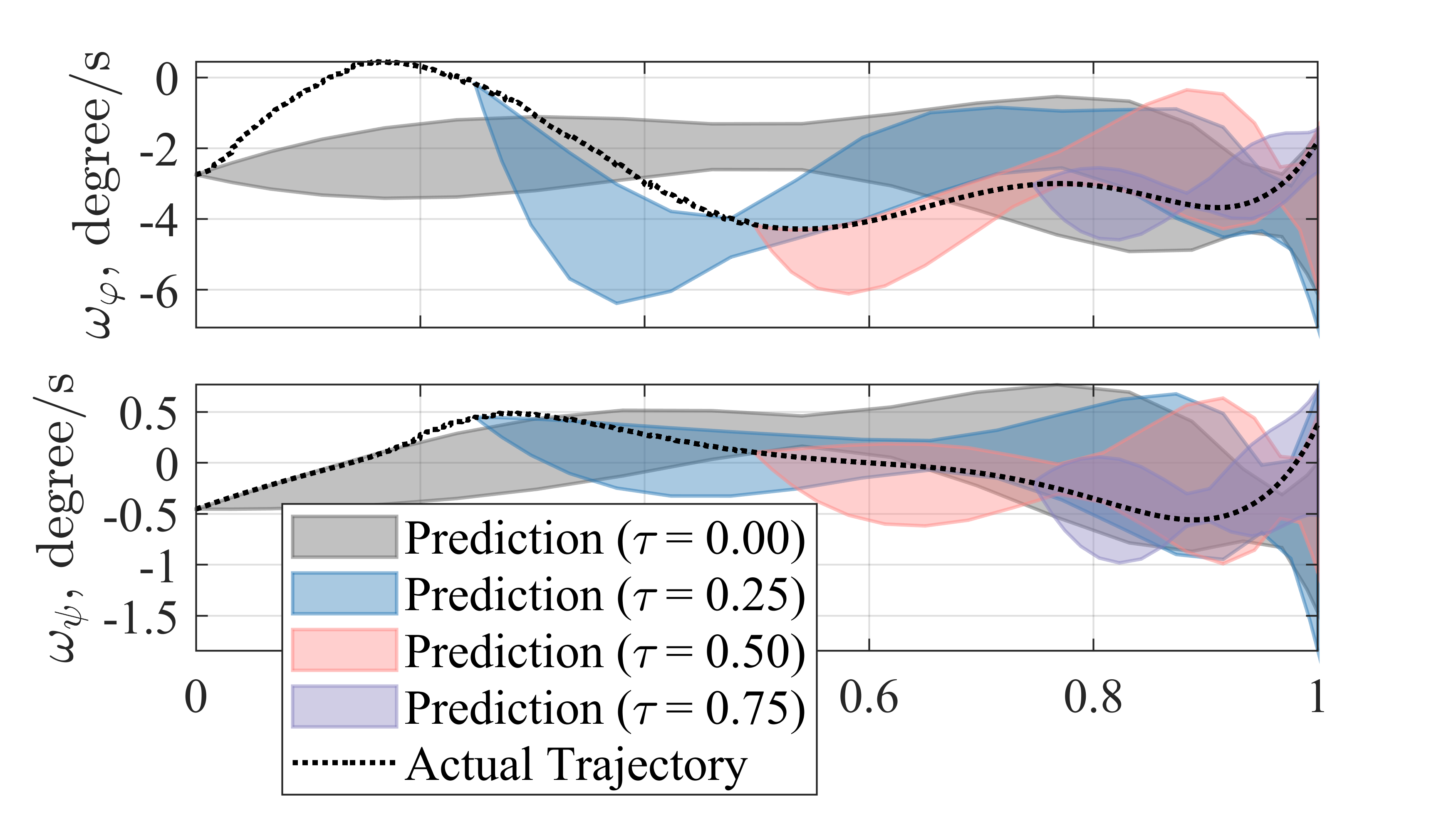}
        \caption{Attitude angular rate trajectory dispersions.}
    \end{subfigure}
    \caption{Online trajectory dispersion prediction results.}
    \label{fig-sim-2}
\end{figure*}

\begin{figure*}[!h]
        \centering
    \begin{subfigure}[h]{0.29\linewidth}
        \centering
        \includegraphics[width=\textwidth]{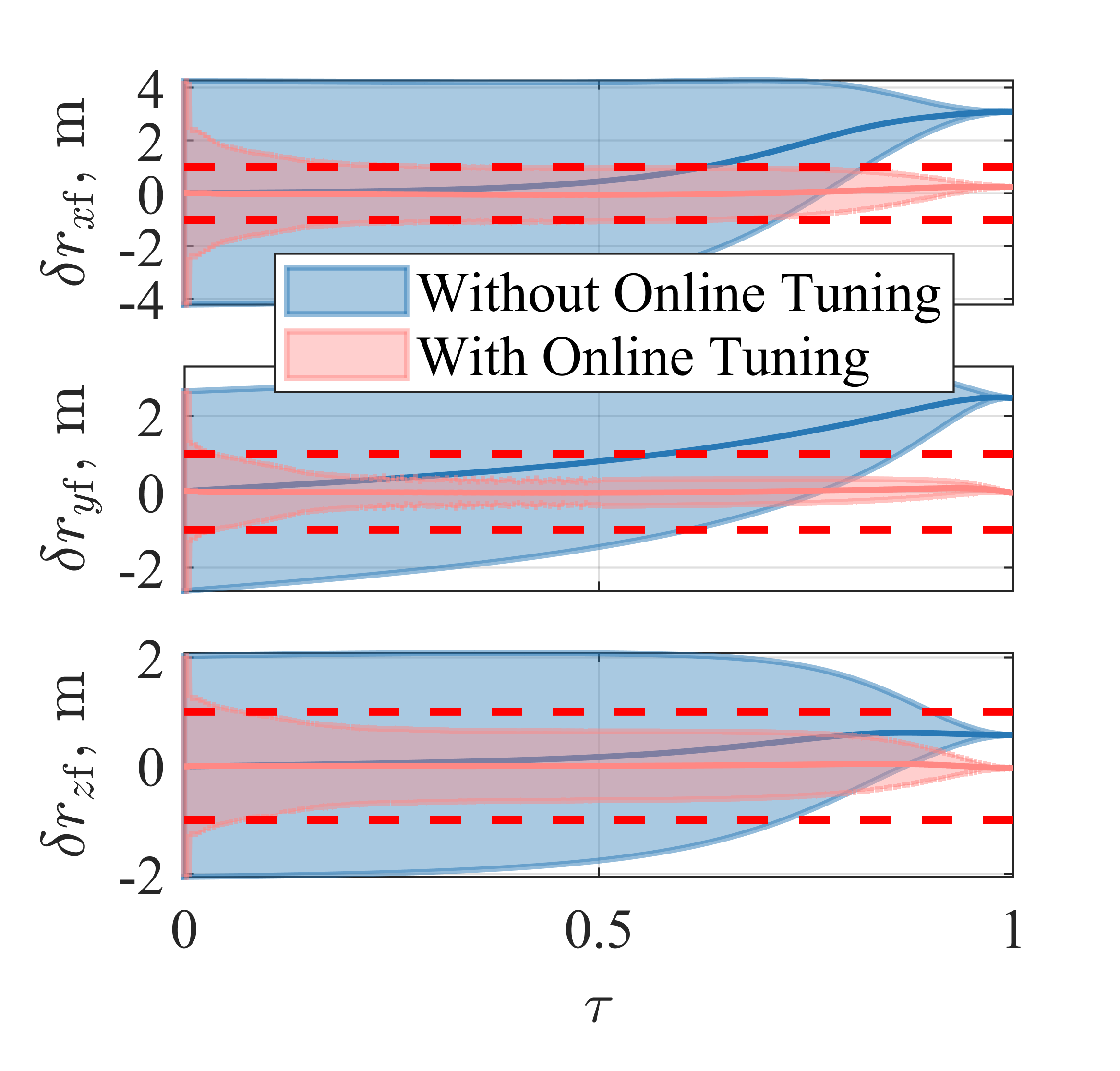}
        \caption{Terminal position dispersions.}
    \end{subfigure}
	 \begin{subfigure}[h]{0.29\linewidth}
        \centering
        \includegraphics[width=\textwidth]{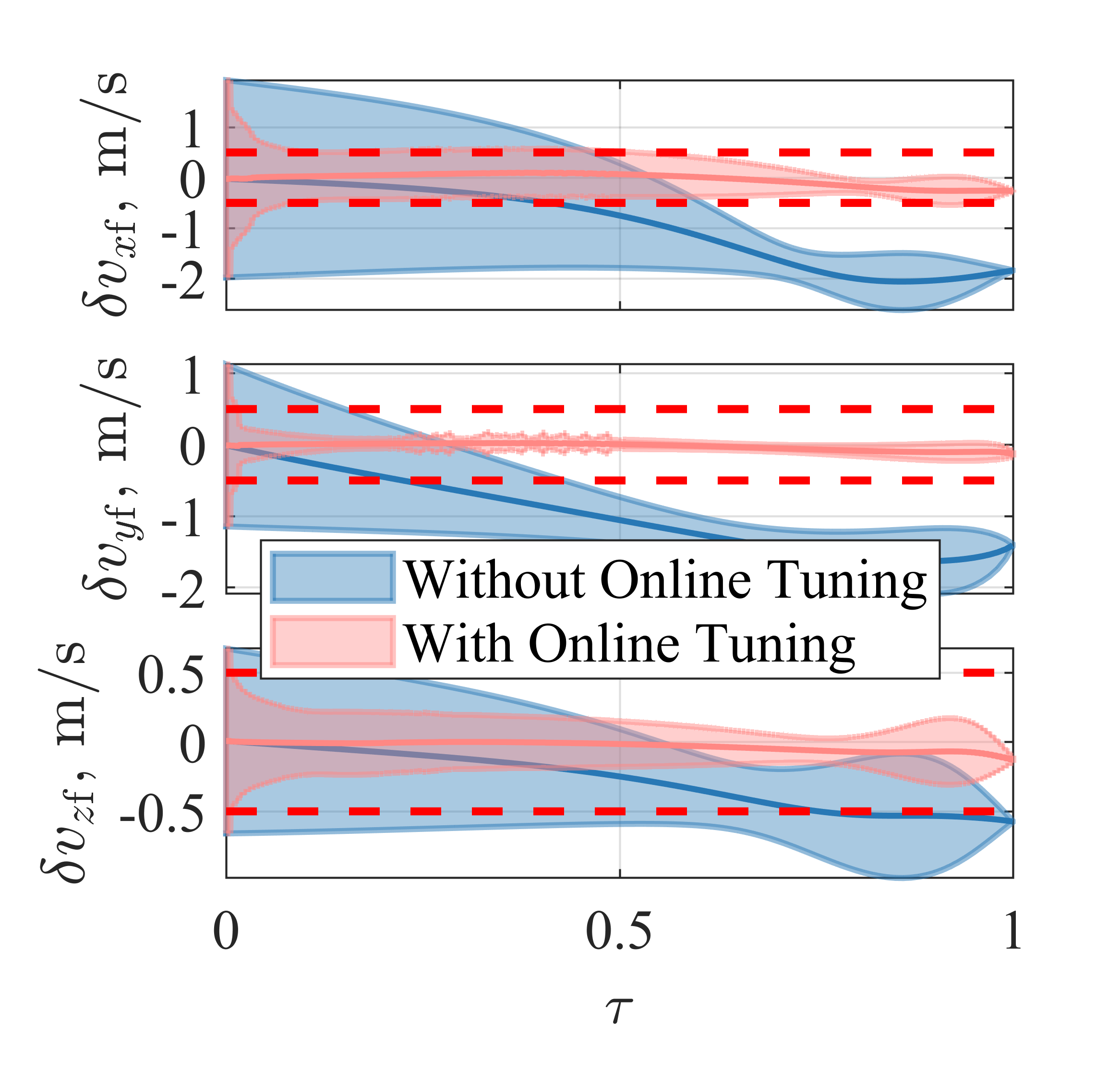}
        \caption{Terminal velocity dispersions.}
    \end{subfigure}
	 \begin{subfigure}[h]{0.29\linewidth}
        \centering
        \includegraphics[width=\textwidth]{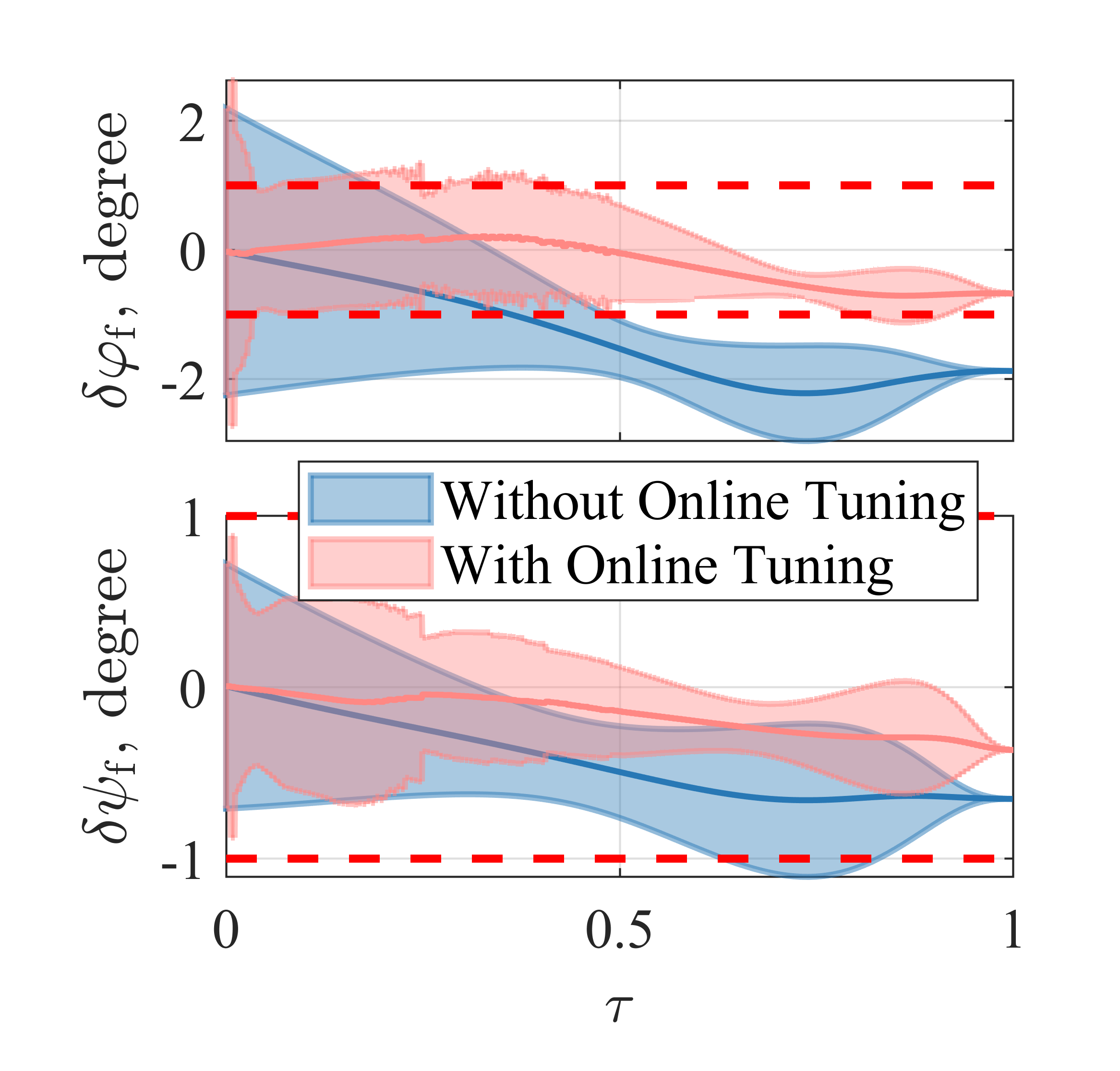}
        \caption{Terminal attitude dispersions.}
    \end{subfigure}
    \caption{Online landing accuracy prediction results.}
    \label{fig-sim-2-Landing}
\end{figure*}

To validate the guidance performance of the proposed trajectory dispersion control method, a simulation using the online parameter tuning is carried out.
The actual disturbances are configured as
\(\xi_{T} = \SI{1.5}{\%}\), \(\xi_{\varphi} = \SI{0.25}{\degree}\), \(\xi_{\psi} = \SI{0.25}{\degree}\), \(\xi_{Cx} = \SI{25}{\%}\), \(\xi_{Cy} = \SI{25}{\%}\), \(\xi_{Cz} = \SI{25}{\%}\), \(\xi_{\rho} = \SI{15}{\%}\), \(\xi_{A} = \SI{45}{\degree}\), \(\xi_{V1} = \SI{15}{m/s}\), \(\xi_{V2} = \SI{30}{m/s}\), and \(\xi_{V3} = \SI{20}{m/s}\).
The actual simulated trajectory and the results of the predicted trajectory dispersion at different moments are shown in \cref{fig-sim-2}.
It can be seen that the trajectory dispersions of position and attitude angle gradually decrease because of the decreasing disturbances in the future.
The guidance command dispersions increase in the initial portion and decrease in the later portion, which is identical with the results in \cref{fig-sim-1-u} and \cref{fig-sim-1-w}.
\cref{fig-sim-2-Landing} gives the landing accuracy prediction results without and with online parameter tuning.
It can be seen that in the case of without online parameter tuning, the terminal landing error can keep within the required range of \(\SI{5}{m}\), \(\SI{2}{m/s}\) and \(\SI{2}{\degree}\), but cannot meet the higher accuracy requirement.
In the case of online parameter tuning, the mean and standard deviation of the terminal landing error dispersion can be rapidly reduced and meet the desired accuracy requirements of \(\SI{1}{m}\), \(\SI{0.5}{m/s}\) and \(\SI{1}{\degree}\).
The above simulation results indicate that the proposed method can adaptively tune the guidance parameter to achieve the trajectory dispersion control and meet the desired landing accuracy.


\section{Conclusion}
\label{Sec-6}

In this note, the trajectory dispersion control method is proposed with two main components: online trajectory dispersion prediction and real-time guidance parameter tuning.
Based on the formulated probabilistic disturbance model, the closed-loop trajectory dispersion is predicted online with high accuracy and is represented as the analytical formulation of the disturbance random vector and the guidance parameter.
By using the simple analytical gradient descent method, the real-time guidance parameter tuning law is designed to achieve the trajectory dispersion control.
Numerical simulations show that the online trajectory dispersion prediction method achieves the same high accuracy as the Monte Carlo method with smaller computational resource;
the real-time guidance parameter tuning law can optimally shape the trajectory dispersion, so that the landing error dispersion is significantly reduced and meets the desired accuracy requirements.
Overall, this note presents a general framework for precision landing guidance:
the POFGL can be easily replaced by other parameterized guidance laws, and the parameter tuning can be achieved using alternative methods, such as Bayesian optimization method and reinforcement learning method.

\bibliography{main}

\end{document}